\documentclass[11pt,oneside]{article}
\usepackage{setspace,graphicx,amssymb,amsmath,latexsym,amsfonts,amscd,amsthm,multirow,ctable,mathdots,caption,array,diagbox,mathtools}
\usepackage{tabularx,cite,mathrsfs}
\usepackage{authblk}

\usepackage[font=footnotesize]{caption}
\usepackage{fullpage}
\usepackage{stmaryrd}
\usepackage{rotating}

\usepackage{hyperref}

\newcommand{\bea}{\begin{eqnarray}}
\newcommand{\eea}{\end{eqnarray}}
\newcommand{\bee}{\begin{eqnarray*}}
\newcommand{\eee}{\end{eqnarray*}}
\newcommand{\al}{\begin{align*}}
\newcommand{\eal}{\end{align*}}
\newcommand{\be}{\begin{equation}}
\newcommand{\ee}{\end{equation}}

\newcommand{\bem}{\begin{pmatrix}}
\newcommand{\eem}{\end{pmatrix}}

\def\a{\alpha}
\def\b{\beta}
\def\c{\gamma}

\def\inf{\infty}

\def\p{\pi}

\def\t{\tau}
\def\th{\theta}
\def\til{\tilde}

\newcolumntype{R}{ >{$}r <{$}}
\newcolumntype{C}{ >{$}c <{$}}
\newcolumntype{L}{ >{$}l <{$}}
\newcolumntype{F}{>{\centering\arraybackslash}m{1.5cm}}

\newcommand{\mc}[1]{\mathcal{#1}}

\newcommand{\comment}[1]{}

\newcommand{\CC}{{\mathbb C}}
\newcommand{\ZZ}{{\mathbb Z}}

\newcommand{\tr}{\operatorname{{tr}}}

\newcommand{\ex}{\operatorname{e}} 

\newcommand{\g}{\gamma}	

\newcommand{\CoN}{Co_1}	

\theoremstyle{definition}

\theoremstyle{remark}

\numberwithin{equation}{section}
\newtheorem*{eg}{Example}

\def\ba#1\ea{\begin{align}#1\end{align}}
\def\bg#1\eg{\begin{gather}#1\end{gather}}
\def\bm#1\em{\begin{multline}#1\end{multline}}
\def\bmd#1\emd{\begin{multlined}#1\end{multlined}}

\begin{document}

\setstretch{1.4}

\title{\vspace{-65pt}
\vspace{20pt}
    \textsc{
    Exceptional Algebra and Sporadic Groups at $c=12$    
    }
}

\author[1]{Miranda C. N. Cheng
}
\author[2]{Sarah M. Harrison
}
\author[3]{Shamit Kachru
}\author[3]{Daniel Whalen
}
\affil[1]{Institute of Physics and Korteweg-de Vries Institute for Mathematics\\\newline
University of Amsterdam, Amsterdam, the Netherlands\footnote{On leave from CNRS, France.}}
\affil[2]{Center for the Fundamental Laws of Nature\\\newline
Harvard University, Cambridge, MA 02138, USA}
\affil[3]{Stanford Institute for Theoretical Physics, Department of Physics\\
and Theory Group, SLAC\\
Stanford University, Stanford, CA 94305, USA}

\date{}

\maketitle

\vspace{-1em}

\abstract{
In earlier works, it was seen that a ${\mathbb Z}/2$ orbifold of the theory of 24 free two-dimensional chiral fermions 
admits various sporadic finite simple groups as global symmetry groups when viewed as an ${\cal N}=1$,
${\cal N}=2$, or ${\cal N}=4$ superconformal field theory.  In this note, we show that viewing
the same theory as an SCFT with extended ${\cal N}=1$ symmetry -- where the extension is
the same one which arises in string compactification on manifolds of exceptional Spin$(7)$ holonomy 
-- yields theories which have global  symmetry given by the sporadic groups $M_{24}, Co_2$ or $Co_3$.  
The  partition functions twined by these symmetries, when decomposed into characters of the Spin(7) algebra,  
give rise to two-component vector-valued mock modular forms encoding an infinite-dimensional module for the corresponding sporadic groups.

}

\clearpage

\tableofcontents


\section{Introduction}
\label{Introduction}

The discovery of `Monstrous Moonshine' \cite{Conway-Norton, FLM} opened a rich  
new area of exploration, uncovering mysterious connections between sporadic simple groups, conformal field theories
(or vertex operator algebras), and string theory.  See, for instance, \cite{gannon} for a comprehensive discussion. 
In recent years, further examples of the moonshine
phenomenon are being uncovered \cite{Eguchi2010,Cheng:2012tq,UMNL}, involving both new objects (such as mock modular forms \cite{Zwegers2008})
and possible connections to further subfields of mathematics (algebraic geometry, for instance).  
For recent reviews with extensive references, see for instance \cite{CheDun_M24MckAutFrms,Gaberdiel:2012um,Duncanreview}.

Here, we contribute a further entry to the list of correspondences between sporadic groups and modular objects.  We view a ${\mathbb Z}/2$ orbifold of the theory of 24 free
chiral fermions in two-dimensions as furnishing an example of the ${\cal SW}(3/2,2)$ superalgebra,
better known as the superalgebra which appears in string compactification on manifolds of exceptional
Spin$(7)$ holonomy \cite{ShVafa}.  The characters of this algebra were studied in \cite{spin7char}, 
building on earlier work of \cite{GepnerNoyvert}.  See \cite{spin7char} for suggestive hints relating this $c=12$ SCFT to sporadic groups.
 In this paper we elaborate on this by describing precise connections between the SCFT
and the groups $M_{24}$, $Co_2$ and $Co_3$.

The organization of this note is as follows.
In \S2, we place this work in a larger context of previous work relating various superalgebras, the present 
$c=12$ SCFT, and various sporadic simple groups.  In \S3, we describe how
the groups $G\cong M_{24}, Co_2$ and $Co_3$ arise as the  global symmetry groups when considering the present SCFT as a representation of the Spin(7) algebra.
In \S4, we briefly review the representation theory of the Spin(7) superalgebra. 
In \S5, 
we first discuss the partition functions 
\begin{equation}
Z_g= {\rm Tr}~g q^{L_0 - {c\over 24} }~
\end{equation}
twined by all elements $g\in G$ of the global symmetry groups mentioned above. 
Subsequently, we discuss the mock modular properties of the 2-vector-valued functions that arise when decomposing $Z_g$ into Spin(7) characters. In \S6 we close the paper with some discussions. 
The appendices include our conventions for standard modular functions (Appendix \ref{sec:modforms}); character tables for the groups which appear (Appendix \ref{sec:characters}); and the coefficient tables (Appendix \ref{app:Coefficient Tables}) together with the corresponding representations 
for the twined partition functions (Appendix \ref{app:decomposition}).

\section{The Free Field Theory}
\label{A Free Field Module}
The two-dimensional $c=12$ chiral conformal field theory that we will consider was first described in \cite{FLM}, and  an alternative construction was given in \cite{Duncan}. 
In \cite{DuncanMack-Crane1}, it was shown that the NS sector of theory is described by a distinguished super vertex operator algebra, $V^{s\natural}$, and can be regarded as the natural analogue of the Frenkel-Lepowsky-Meurman moonshine module for Conway's largest group. In this section we describe the two constructions of this 2d chiral conformal field theory, and review how its NS-sector twined partition functions are given by certain normalized principal moduli with vanishing constant terms. 

The construction presented in \cite{FLM} contains 8 free bosons $X^i$ compactified on the eight-dimensional torus ${\mathbb R}^8/\Lambda_{E_8}$ given by the $E_8$ lattice plus their 8 fermionic superpartners $\psi^i$, subject to a $\mathbb Z/2$ orbifold
\begin{equation}
(X^i,\psi^i) \to (-X^i,-\psi^i)~.
\end{equation}
Clearly, this theory exhibits a manifest $\mathcal N=1$ supersymmetry. 
In addition, the orbifolding ensures that there are no NS primary fields of dimension ${1\over 2}$ \footnote{Here and in the rest of the paper we use the common terminology and refer to the conformal weight of a state, or the $L_0$-eigenvalue of an eigenstate, as ``dimension". }. The partition function of the theory in the NS sector is given by
\begin{eqnarray}
\label{E8Z}
Z_{NS,E8}(\t) &=& \tr_{NS}q^{L_0-c/24}= {1\over 2} \left( { {E_4\theta_3^4} \over \eta^{12}} + 16 {\theta_4^4 \over \theta_2^4} +
16 {\theta_2^4 \over \theta_4^4} \right)(\tau)~\\
&=&  {q}^{-1/2} + 0 + 276\, {q}^{1/2} + 2048\, q + 11202 \, q^{3/2} + \cdots,~
\end{eqnarray}
where we write $q=e(\tau)$ and use the shorthand notation $e(x):=e^{2\p i x}$. 
In the above
$E_4(\tau)$ is the weight 4 Eisenstein series defined in (\ref{def:E4}), $\eta(\t) = q^{1/24}\prod_{n=1}^\inf (1-q^n)$ is the Dedekind eta function, and $\th_i(\t)$ are the Jacobi theta functions recorded in Appendix \ref{sec:modforms}. The authors of \cite{FLM} observed that these coefficients of this function are related to representations of the sporadic group $Co_1$. Apart from 276 which simply coincides with the dimension of an irreducible representation of ${{Co}}_1$, one can also observe
\begin{eqnarray}
2048 & =& 1 + 276 + 1771\,,\\
11202 &=& 1 + 276 + 299 + 1771 + 8855\,
\end{eqnarray}
and so on.

Another construction of this module which renders its symmetries manifest was discussed in \cite{Duncan}. This module is built from 24 free chiral fermions, $\lambda_\alpha, ~\alpha=1\ldots 24$ with again an  orbifold by the order 2 symmetry $\lambda_\alpha\to -\lambda_\alpha$. From this point of view, the partition function is more naturally written as
\begin{equation}
\label{Zfermion}
Z_{NS,\rm fermion}(\t) = {1\over 2} \sum_{i=2}^{4} {\theta_i^{12}(\t,0) \over \eta^{12}(\t)}~,
\end{equation}
which is also equivalent to the following infinite product expression:
\be
Z_{NS,\rm fermion}(\t) = q^{-{1\over 2}}\prod_{n=1}^\infty (1+q^{n-{1\over 2}})^{24}-24.
\ee
Note that in (\ref{Zfermion})  and similarly in the twining function (\ref{def:twining}) below one could in principle extend the sum over  ${\theta_i^{12}(\t,0) \over \eta^{12}(\t)}$ to $i=1,\dots,4$, but this does not make a difference since $\theta_1^{}(\t,0)=0$. 
As a theory of 24 free chiral fermions, it is apparent that the symmetry group of this model is Spin(24). Moreover, in \cite{Duncan} it is shown that by choosing a particular linear combination of dimension-${3\over 2}$ spin fields, one can construct an $\mathcal N=1$ super-current. This choice of the supercharge breaks the Spin(24) symmetry to $\CoN$. Therefore, as an $\mathcal N=1$ module this theory has a discrete global symmetry group $\CoN$.

Given this symmetry group, one can compute the partition function twined by any $g\in \CoN$ as follows. Define $\epsilon_a,~a=1\ldots 24$ to be the eigenvalues of $g$ in the 24-dimensional non-trivial representation \textbf{24} of $\CoN$, satisfying $\epsilon_a =\overline{\epsilon_{a+12}}$ and ${\rm Im}(\epsilon_a )\geq 0$, and  define also $z_i\pmod{\mathbb Z}$ by $e(z_a)=\epsilon_a$ and $z_i\in [0,1/2]$ . The partition function of the theory twined by $g$ is then given by:
\be\label{def:twining}
Z_{NS,g}(\t)=\tr_{NS} g q^{L_0-c/24} =  {1\over 2} \sum_{i=2}^{4} \varepsilon_i(g) \prod_{a=1}^{12} {\theta_i(\t,z_a ) \over \eta(\t)}=q^{-{1\over 2}}\prod_{n=1}^\infty \prod_{a=1}^{24}(1+\epsilon_a q^{n-{1\over 2}})-\chi_g,
\ee 
where $\chi_g=\tr_{\rm \textbf {24}}g =\sum_{a=1}^{24} \epsilon_a$. 
In the above we have defined $\varepsilon_i(g)$ to encode the action of $\langle g\rangle$ on the $2^{12}$-dimensional spinor representation, by setting
\be
\varepsilon_2(g) =\frac{{\rm Tr}_{\bf 4096} g}{2^{12}\prod_{a=1}^{12} \cos(\pi z_a) }\in \{-1,1\}~,~~ \varepsilon_3(g)=\varepsilon_4(g)=1. 
\ee
See Appendix \ref{sec:Frame Shapes} for the value of ${\rm Tr}_{\bf 4096} g$ for $g\in G$ and $G\cong M_{24}, Co_2, Co_3$ and see  Lemma 4.8 and Theorem 4.9 of \cite{DuncanMack-Crane1} for a proof of the last equality in (\ref{def:twining}). 

In fact, in \cite{DuncanMack-Crane1} it was shown that a closely related set of functions, given by
\be
T'_g(\t)=\tr_{NS}(-1)^F g q^{L_0-c/24}= - Z_{NS,g}(\t+1) 
\ee
have the property that $T_g(\t)= T'_g(2\t)$ is the normalized principal moduli for certain genus zero group $\Gamma_g$ for every $g \in \CoN$. (Clearly, this implies that $ Z_{NS,g}(\t)$ is a principal modulus for the corresponding group related to $\Gamma_g$ by a conjugation.) 
We say that a group $\Gamma \subset SL_2(\mathbb R)$ is genus zero if $\Gamma\backslash\hat {\mathbb H}$ is isomorphic to a Riemann sphere, where $\hat{\mathbb H}:={\mathbb H}\cup {\mathbb Q}\cup \infty$ denotes the compactified upper half plane. A function invariant under $\Gamma$ is said to be a   principal modulus for $\Gamma$ if it is given by a isomorphism $\Gamma\backslash\hat {\mathbb H} \to {\mathbb C\mathbb P}^1$, and it is said to be a normalised principal modulus  if it moreover has the expansion $q^{-1}+O(q)$ in the limit $\tau\to i\infty$.

In \cite{M5} it was shown that one can extend this construction by building larger superalgebras out of the fermions. A choice of two (three) fermions is required to generate the $U(1)$ ($SU(2)$)  R-symmetry currents of an $\mathcal N=2$ ($\mathcal N=4$) superalgebra. This breaks the symmetry group of the theory to a subgroup of $\CoN$ which stabilizes a two-(three-)plane in the Leech lattice and by construction commutes with the corresponding superalgebra. 
Consequently the chiral CFT, when viewed as  $\mathcal N=2$ ($\mathcal N=4$) superconformal field theory, results in modules for subgroups $G'$ of $\CoN$ that stabilize two-  (three-)planes in the Leech lattice. 
In other words, decomposing the ($U(1)$-graded) partition function twined by $g\in G'$ into characters of the corresponding superalgebra yields a two-component vector-valued mock modular form whose coefficients are given by certain $g$-characters. 
It was also established in \cite{M5} that the groups $M_{23}$ and $M_{22}$ are distinguished among the groups which preserve an $\mathcal N=2$ or $\mathcal N=4$ algebra, respectively.  This is because all of their twining functions satisfy a certain ``optimal pole" condition, and are thus candidates for the moonshine phenomenon.

In the rest of this paper, we explore a similar story for an extended $\mathcal N=1$ superalgebra, the algebra which, as discussed in \cite{ShVafa}, is associated with $c=12$ conformal field theories with target manifolds of exceptional holonomy Spin(7). This algebra, which can be defined as an $\mathcal N=1$ superalgebra extended by a single copy of the Ising model, breaks the global symmetry group of the Conway module from $\CoN$ to a subgroup which stabilizes a line in the Leech lattice. The choice of line corresponds to the choice of single fermion used to generate the $c={1\over 2}$ Ising sector of the extended $\mathcal N=1$. We compute twining functions for a number of groups which preserve such a ``1-plane" in the Leech lattice and show they are given in terms of two-component vector-valued mock modular forms.  
 See Table \ref{tbl:groups} for a summary of the extended superalgebras in the Conway module and examples of the groups they preserve. 
The Mathieu groups $M_{22}$, $M_{23}$ and $M_{24}$ are singled out here as the 3-, 2-, 1-plane preserving groups in the Leech lattice that admit a representation as permutation groups on 24 objects. 
See \cite{sphere_packing} for more details on these groups and their relation to the Leech lattice. 

\begin{table}[htb]
\begin{center}
\begin{tabular}{c|c|c}
Superalgebra&  Geometrical Representation & Global symmetry group\\\hline
$\mathcal N=0$& ${\mathbb R}^{24}$& $Spin(24)$\\
$\mathcal N=1$&$\Lambda_{Leech}$ & $\CoN$\\
Spin(7)&$\Lambda_{Leech}$, fixed 1-plane & $M_{24}$\\
$\mathcal N=2$&$\Lambda_{Leech}$, fixed 2-plane & $M_{23}$\\
$\mathcal N=4$&$\Lambda_{Leech}$, fixed 3-plane & $M_{22}$\\
\end{tabular}\caption{The extended superconformal algebras compatible with the chiral CFT, some of the possible resulting unbroken global symmetry groups, and the relevant geometric objects these symmetry groups act on.  
}\label{tbl:groups}
\end{center}
\end{table}

\section{Exceptional Holonomy and Sporadic Groups}
Given an $\mathcal N=1$ superconformal algebra and a single fermion, in \cite{ShVafa} it was shown that one can generate the extended $\mathcal N=1$ superconformal algebra associated with manifolds of exceptional holonomy Spin(7). See \cite{ShVafa} or \cite{spin7char} for more details of this construction, as well as the precise definition of this algebra. Given a choice of such an extended $\mathcal N=1$ algebra in the  SCFT in the NS sector, we would like to know the global symmetry group leaving such a choice invariant. Recall that constructing an $\mathcal N=1$ supercurrent in the free fermion orbifold theory breaks the global symmetry group from Spin(24) to $\CoN$, while $Co_0\cong \ZZ/2.Co_1$ is the automorphism group of $\Lambda_{Leech}$. 
In chapter 10 of \cite{sphere_packing} it is explained how certain subgroups of $Co_0$ can be described as stabilizers of certain  lattice vectors in $\Lambda_{Leech}$. 
Given an $\mathcal N=1$ SCA, one needs a single fermion, and thus a choice of $\mathbb R\in \mathbb R^{24}$, to enhance it to the Spin(7) algebra. The choice of $\mathbb R$ corresponds to the choice of fermion used to construct the Ising sector of the superalgebra. Therefore, the corresponding global symmetry group which commutes with this algebra must correspond to a subgroup of $Co_1$ which further stabilizes a line in the Leech lattice. We will discuss three such sporadic simple groups, each of which stabilizes a different class of vector in the Leech lattice \cite{sphere_packing}:
\begin{itemize}
\item The Mathieu group $M_{24}$, of order $2^{10}\cdot3^3\cdot5\cdot7\cdot11\cdot23~(= 244823040)$
\item The group $Co_3$ of order $2^{10}\cdot3^7\cdot5^3\cdot7\cdot11\cdot23~(=495766656000)$
\item The group $Co_2$ of order $2^{18}\cdot3^6\cdot5^3\cdot7\cdot11\cdot23~(=42305421312000)$
\end{itemize}
See the tables in Appendix \S\ref{sec:characters} for the characters of their irreducible representations. 

With similar reasoning to that in \cite{M5}, a choice of a one-dimensional subspace in the $\textbf{24}$ of $Co_0$, {\it i.e.} a choice of single linear combination of the 24 fermions, corresponds to a specific copy of an extended $\mathcal N=1$ superconformal algebra and a specific stabilizing subgroup of $Co_0$. This amounts to a proof that the chiral CFT leads to an infinite-dimensional $G$-module for each of $G\in \{M_{24}, Co_2,Co_3\}$ when viewed as an extended $\mathcal N=1$ module. In the following we will show this explicitly, and we will find that the graded characters of the corresponding $G$-modules are naturally given in terms of mock modular forms. 

Note that when the a 1-plane  ${\cal M}$ lies inside a 2-(3-)plane  ${\cal M}'$ in the Leech lattice, the corresponding 
extended $\mathcal N=1$ superconformal algebra ${\cal A}({\cal M})$ is contained in an $\mathcal N=2$ ($\mathcal N=4$) superconformal algebra ${\cal A}({\cal M}')$ corresponding to a 2-(3-)plane  ${\cal M}'$, and the global symmetry $G_{{\cal A}({\cal M})}$ of the chiral CFT viewed as an ${\cal A}({\cal M})$-module naturally contains  the global symmetry $G_{{\cal A}({\cal M}')}$of the same chiral CFT viewed as an ${\cal A}({\cal M}')$-module. Namely, 
\[
{\cal M} \subset {\cal M}' \Rightarrow {\cal A}({\cal M})\subset  {\cal A}({\cal M}')~,~ G_{{\cal A}({\cal M})}\supset G_{{\cal A}({\cal M}')}. 
\]
For instance, comparing to the groups arising in the analysis in \cite{M5} we have $M_{22} \subset M_{23}\subset M_{24}$, and both the McLaughlin group ($McL$) and the  Higman-Sims group ($HS$) are  subgroups of both $Co_2$ and $Co_3$.

\section{Characters of the Spin(7) algebra}
In this section we briefly review the results of the computation of characters of the Spin(7) algebra from \cite{spin7char} and their modular properties. 
The unitary representations of the $SW(3/2,2)$ superconformal algebra were classified in \cite{GepnerNoyvert}. The algebra for manifolds with holonomy Spin(7) discussed in \cite{ShVafa} is a special case of this algebra with $c=12$.  At $c=12$ the algebra can be thought of as an extension of the $\mathcal N=1$ superconformal algebra by two additional generators (the stress-energy tensor of the Ising model, and its superpartner), with the constraint that the algebra closes. The algebra can be realized in two ways corresponding to whether the fermionic operators are integer or half-integer graded. These are respectively the Ramond and the Neveu-Schwarz Spin(7) algebra, which are isomorphic to each other.

The unitary highest weight states in the NS sector can be specified by two quantum numbers: the dimension of the internal Ising factor, $a$, and the total dimension, $h$. There are three massless representations and two continuous families of massive representations. In the Ramond sector, the massive states have two components and are labeled by Ising dimensions $(a_1,a_2)$ and total dimension $h$. 
The NS and Ramond sector states are in one-to-one correspondence via the relation given in Table \ref{tbl:NSR},  where states are labeled by $|a,h\rangle$ and the correspondence is between the states in the same row.

\begin{table}[htb]\begin{center}
\begin{tabular}{ccc}
NS& &R\\\hline
$\left |0,0\right\rangle$ &$  $& $\left |{1\over 2},{1\over 2}\right \rangle$\\
$\left |\frac{1}{16},{1\over 2}\right\rangle$ &$  $& $\left |{1\over 16},{1\over 2}\right \rangle$\\
$\left |{1\over 2},1\right\rangle$ &$ $ & $\left |0,{1\over 2}\right \rangle$\\
$\left |0, n\right\rangle$ &$ $ & $\left |\left ({1\over 16},{1\over 2}\right ),{1\over 2}+n\right \rangle$\\
$\left |{1\over 16},{1\over 2} + n\right\rangle$ &$ $ & $\left |\left (0,{1\over 16}\right ),{1\over 2}+n\right \rangle$\\
\end{tabular}\caption{The unitary irreducible highest weight representations of the Spin(7) algebra, where $n\in\mathbb Z_{> 0}$. The NS and Ramond sector states are in one-to-one correspondence by relating the states in the same row. The states in the first three rows are the so-called massless states.}\label{tbl:NSR}\end{center}
\end{table}

The characters for each of these states were conjectured in \cite{spin7char}, and are given in terms of standard (mock) modular forms. Here we briefly review these results; see \cite{spin7char} for more details and derivations. First define, for all positive integers $m$ and  $r\in \ZZ/2m\ZZ$,
 \be
\theta_{m,r}(\tau)  = \sum_{k =r\!\pmod{2m}} q^{\frac{k^2}{4m}}.
\ee
Note that $\theta_{m,r}(\t)= \theta_{m,-r}(\t)$ and the vector-valued function $\th_{m, r}(\t)$ transforms under $SL_2(\ZZ)$ as
\begin{align}
\theta_{m}(-\frac{1}{\t}) &= \sqrt{-i \t}\,  {\cal S}^{(\th)} \,\theta_{m}(\t,z),\qquad\label{transf_theta}\\
\theta_{m}({\t}+1,{z}) &= {\cal T}^{(\th)} \,\theta_{m}(\t,z),
\end{align}
where ${\cal S}$, ${\cal T}$ are the $2m \times 2m$ unitary matrices with entries
\begin{align}
{\cal S}^{(\th)}_{rr'} &= \frac{1}{\sqrt{2m}} e^{\frac{\pi i r r'}{m}},\\ \label{transf_theta_2}
{\cal T}^{(\th)}_{rr'} &= e^{\frac{\pi i r^2}{2m}} \delta_{r,r'}.
\end{align}

We also define
\be\label{eq:AL}
f_u^{(m)}(\tau,z)= \sum_{k\in \mathbb Z} \frac{q^{mk^2} y^{2mk}}{1- yq^k e^{-2\pi iu}}.
\ee
Note that   $f_u^{(m)}$ has the so-called elliptic transformation property: 
\[
f_u^{(m)}(\tau,z)  = f_u^{(m)}(\tau,z+1) = q^m y^{2m} f_u^{(m)}(\tau,z+\t) .
\]
Its completion, defined as \cite{Zwegers2008}
\be\label{def:completion_f}
\hat f_u^{(m)}(\tau,\bar \tau,z)  = f_u^{(m)}(\tau,z)-{1\over2}\sum_{r\in\ZZ/\mathbb Z}R_{m,r}(\tau,u)\theta_{m,r}(\tau,z) 
\ee
with 
\begin{align}\notag
R_{m,r}(\tau,u)&=\sum_{k=r\!\!\pmod{2m}}\left({\rm sgn}(k+\tfrac{1}{2}) -E(\sqrt{{\rm Im}\t\over m} (k+2m\frac{{\rm Im}u}{{\rm Im}\t})) \right)
 q^{-k^2\over 4m}e^{-2\pi i k u} \\ \notag
 E(z) &= {\rm sgn}(z) \left(1-\int_{z^2}^\infty dt\,t^{-1/2} \,e^{-\pi t}\right),
\end{align}
transforms as a Jacobi form of weight 1 and index $m$. 
For later convenience we also define 
\be
\til \th_{m,r} (\t)= \th_{m,r} (\t) + \th_{m,r-m} (\t)  
\ee
satisfying $\til \th_{m,r}=\til \th_{m,-r}=\til \th_{m,r+m}$, $\til \th_{m,r}(\tau) = \theta_{m/2,r}(\tau/2)$, and
\be
\tilde f_u^{(m)}(\tau,z)=f_u^{(m)}(\tau,z)-f_u^{(m)}(\tau,-z).
\ee

The characters $\chi^{NS}_{a,h}$  for the massive states with dimensions $|a,h\rangle$ in the NS sector are given as
\begin{align}
\chi^{NS}_{0,h}(\t)&=q^{h-{49\over 120}}\, \mc P(\t)\, \Theta^{NS}_{0}(\t)
= q^{h}\, (q^{-1/2}+1 +q^{1/2}+3\,q +\dots)
\end{align}
and
\be
\chi^{NS}_{{1\over 16},h}(\t)=q^{h-{61\over 120}}\, \mc P(\t)\, \Theta^{NS}_{1\over16}(\t)
=  q^{h} \,(q^{-1/2}+2 +3\,q^{1/2}+5\,q +\dots)
\ee
where 
\begin{align}
\Theta^{NS}_{0}(\t)&= \Big(\til\th_{30,2}(\t)-\til\th_{30,8}(\t)\Big) = \sum_{k\in \ZZ}\epsilon^{NS}_{0}(k) \,q^{\frac{k^2}{120}}\\
\Theta^{NS}_{{1\over 16}}(\t)&=  \Big(\til\th_{30,4}(\t)-\til\th_{30,14}(\t)\Big) = \sum_{k\in \ZZ}\epsilon^{NS}_{{1\over 16}}(k) \,q^{\frac{k^2}{120}}\\
\epsilon^{NS}_{0}(k)& = \begin{cases}1&k= 2,28\!\!\! \pmod{60}\\-1&k=- 8,-22\!\!\! \pmod{60}\\0&\text{otherwise}\end{cases}\\
\epsilon^{NS}_{{1\over 16}}(k)& = \begin{cases}1&k= 4,26\!\!\! \pmod{60}\\-1&k= -14,-16\!\!\! \pmod{60}\\0&\text{otherwise}\end{cases}
\end{align}
and we have defined
$$\mc P(\t)=\frac{\eta^2(\tau)}{\eta^2 (\tfrac{\t}{2}) \eta^2(2\tau)}.$$ 
The massless character of total dimension $h={1\over 2}$ is given by 
\be
\tilde{\chi}^{NS}_{1\over2}=  \mc P(\t)\,\mu^{NS}(\tau)~.
\ee 
Here,
\be\label{def:mu}
\mu^{NS}(\tau)= \big( q^{5\over 8} 
\tilde f^{(5)}_{{\tau\over 2} + {1\over 2}}(6\tau,\tau)+q^{25\over 8}\tilde f^{(5)}_{{\tau\over 2} + {1\over 2}}(6\tau,-2\tau)\big),
\ee
and the other two massless characters can be found using the BPS relations which relate massless and massive characters:
\be\label{rel_massless_massive1}
\tilde{\chi}^{NS}_0 + \tilde{\chi}^{NS}_{1\over16}= q^{-n}\chi^{NS}_{0,n}
\ee
and
\be\label{rel_massless_massive2}
\tilde{\chi}^{NS}_{1\over16} + \tilde{\chi}^{NS}_{1\over2}= q^{-n}\chi^{NS}_{{1\over16},{1\over2}+n},
\ee
where we use $\tilde{\chi}^{NS}_a$ in the above to denote the character of the massless state of Ising dimension $a$ in the NS sector.

Consider the function $\mu^{NS}$ defined in (\ref{def:mu})
appearing in  formula for the massless character $\tilde{\chi}^{NS}_{1\over2}$. 
From its definition and (\ref{def:completion_f}) we see that 
one can define a (non-holomorphic) completion of this function which transforms as a weight $1$ modular form under $\Gamma_\theta$. 
Using the identity 
\[
R_{m,r}(\tau,u=\a \tau+ \b) = {1\over \sqrt{2m}} \,q^{m\a^2}\,e(-r\beta-\tfrac{1}{8})\,\int_{-\bar \tau}^\infty d\t' \, (\t'+\t)^{-1/2} \left( \sum_{k=r\pmod{2m}} (k+2m\a) \,e^{2\pi i \tau' {(k+2m\a)^2\over 4m}} \right), 
\]
for $0<2m\alpha< 1$, $2m\beta\in\ZZ$, 
we see that the completion is given by 
\be
\hat{ \mu}^{NS}(\tau,\overline \tau)=\mu^{NS}(\tau)-{1\over 2}\frac{1}{\sqrt{60 i}}\int_{-\overline \tau}^{i\infty} d\tau'~(\tau' + \tau)^{-{1\over 2}} \underline{\theta}_{NS}(\t) \cdot  \underline{S}(\tau')
\ee
where we have defined
\begin{align} \underline{\theta}_{NS}=\left( \begin{array}{c}
\Theta^{NS}_{1\over 16} \\
\Theta^{NS}_0 \end{array} \right), ~
 \underline{S}=\bem S_1 \\ S_7\eem
 \end{align}
 Here, 
 $S_\alpha(\tau)= \sum_{k\in\mathbb Z}  k \epsilon^{R}_{\alpha}(k) q^{k^2/120} 
 $ for $\alpha=1,7$ and 
 \begin{align}
 \epsilon^{R}_{1}(k)& = \begin{cases}1&k= 1,29\!\!\! \pmod{60}\\-1&k= -11,-19\!\!\! \pmod{60}\\0&\text{otherwise}\end{cases}\\
\epsilon^{R}_{7}(k)& = \begin{cases}1&k= -7, -23\!\!\! \pmod{60}\\-1&k= 17,13\!\!\! \pmod{60}\\0&\text{otherwise}\end{cases}.
 \end{align}

Including the factor of $\mc P$, the character $\tilde{\chi}^{NS}_{1\over2}=\mc P \mu^{NS}$ as a whole transforms as a weight $0$ mock modular form under~$\Gamma_\theta$, defined as 
\be
\Gamma_\theta = \left\{ \begin{pmatrix} a & b\\ c&d\end{pmatrix}\in SL_2(\ZZ) \Big\lvert\, c-d \equiv a-b \equiv 1\pmod{2} \right\}.
\label{eq:gammatheta}
\ee

The characters in the Ramond sector take a very similar form. 
The massive ones are  given by
\begin{align}\notag
\chi^{R}_{\left(0,{1\over 16}\right ),h+{1\over2}}(\t)&=2\, q^{h-{1\over 120}} \frac{\eta(2\tau)^2}{\eta(\tau)^4}  \Theta^{R}_{(0,{1\over 16} )}(\t)
= 2\,q^{h}\, (1+3q+8q^2+19q^3+\dots)
\end{align}
and
\be
\chi_{\left ({1\over 16},{1\over 2}\right ),h+{1\over2}}^R(\t)=2 \, q^{h-{49\over 120}} \frac{\eta(2\tau)^2}{\eta(\tau)^4}    \Theta^{R}_{({1\over 16},{1\over2})}(\t)
=  2\,q^{h} \,(1+3q+7q^2+ 16q^3+\dots)
\ee
where 
\begin{align}
 \Theta^{R}_{(0,{1\over 16} )}(\t)&= \Big(\til\th_{30,1}(\t)-\til\th_{30,11}(\t)\Big) = \sum_{k\in \ZZ}\epsilon^{R}_{1}(k) \,q^{\frac{k^2}{120}}\\
 \Theta^{R}_{({1\over 16},{1\over2})}(\t) &=  \Big(\til\th_{30,7}(\t)-\til\th_{30,17}(\t)\Big) = \sum_{k\in \ZZ}\epsilon^{R}_{7}(k) \,q^{\frac{k^2}{120}}
\end{align}
We refer to \cite{spin7char} for a more detailed discussion on the Ramond sector characters.

\section{The Modules}
In \cite{spin7char} it was shown that the graded partition function $Z_{NS}(\t)$ of equation (\ref{Zfermion}) has a decomposition into characters of the Spin(7) algebra of the form
\be
Z_{NS}= a_0 \tilde{\chi}^{NS}_0 + a_{1\over 16} \tilde{\chi}^{NS}_{1\over16} + a_{1\over2}\tilde{\chi}^{NS}_{1\over2} + \sum_{n=1}^\infty b_n \chi^{NS}_{0,n} + \sum_{n=1}^\infty c_n\chi^{NS}_{{1\over 16}, {1\over 2}+n}
\ee
where $a_0, a_{1\over16}, a_{1\over2}, b_n$, and $c_n$ are all non-negative, integral constants. 

Note that it is necessary to also use  the Ramond-sector partition function and the correspondence between NS and R sector states, summarised in Table \ref{tbl:NSR}, to fix these constants.
 The result is $a_0=1, a_{1\over16}=0, a_{1\over 2}=23$ for the massless multiplets, and the first few coefficients for the massive states are given as follows:
\be
b_1=253, b_2=7359, b_3=95128, \ldots
\ee
and
\be
c_1=1771, c_2=35650, c_3=374141, \ldots~.
\ee 
Using the relations between the massless and massive characters given in (\ref{rel_massless_massive1})-(\ref{rel_massless_massive2}), we can repackage the decomposition in the following form
\be\label{eq:ZNS}
Z_{NS}= {\mathcal P}\left ( 24 \mu^{NS} + f_1 \Theta^{NS}_{1\over 16}+ f_7\Theta^{NS}_{0} \right),
\ee
where
\begin{align}\label{fone}
\nonumber f_1(\tau)&=q^{-{1\over 120}}(-1 + c_1 q + c_2 q^2 + c_3 q^3 + \ldots) \\
&=q^{-{1\over 120}}(-1 + 1771 q + 35650 q^2 + 374141 q^3 + \ldots)
\end{align}
and 
\begin{align}\label{fseven}
\nonumber f_7(\tau)&= q^{-{49\over 120}}(1 + b_1 q + b_2 q^2 + b_3 q^3 + \ldots),\\
&=q^{-{49\over 120}}(1 + 253 q + 7359 q^2 + 95128 q^3 + \ldots).
\end{align}

Note that this decomposition procedure entirely analogous to the corresponding $\mathcal N=2$ and $\mathcal N=4$ decompositions of the present CFT
\cite{M5}. 

The modular properties of $\mu^{NS}$ and $\underline{\theta}_{NS}$ given in the previous section imply that $\underline f :=  (\begin{smallmatrix}f_1 \\ f_7\end{smallmatrix})$ is as a weakly holomorphic weight-${1/ 2}$ vector-valued mock modular form for $SL_2(\mathbb Z)$ with shadow $24\underline S(\t)$.
From (\ref{transf_theta})-(\ref{transf_theta_2}) it is straightforward to see that $\underline{S}$ is a vector-valued modular form of weight 3/2 with the multiplier system $\sigma:SL_2(\ZZ)\to GL_2(\CC)$: 
\be\label{multi_shadow}
(c\t+d)^{-3/2}\,\sigma(\g) \,\underline{S}(\tfrac{a\t+b}{c\t+d})  =  \underline{S}(\t) \quad,\quad {\rm for~all}~~\gamma=\bem a&b\\c&d \eem  \in SL_2(\ZZ). 
\ee
Explicitly, $\sigma$ satisfies 
\begin{align} 
\underline{S}(-{1\over\tau}) &=  \tau^{3/2} \,{\cal S}_{} \underline{S}(\t) \\
\underline{S}(\t+1) &={\cal T}_{}   \underline{S}(\t),
\end{align}
where 
\begin{align}
{\cal S}_{} & = \ex({1\over 8}) \bem- \a& \b\\ \b&  \a\eem\\
{\cal T}_{} &= \bem \ex(\tfrac{1}{120})& 0\\ 0&\ex(\tfrac{49}{120})\eem 
\end{align}
and
\begin{align}\notag
\alpha  =\sqrt{\frac{2}{5+\sqrt{5}}} ~,~~
\beta  = \sqrt{\frac{2}{5-\sqrt{5}}} .
\end{align}

Recall that we have shown that this theory furnishes a module for the Spin(7) superconformal algebra admits an action of a group $G$ whenever $G$ stabilizes a line in the Leech lattice. Therefore,  the twined partition function $Z_{NS,g}$ for $g\in G$ admits a decomposition of the form
\be\label{eq:Zg}
Z_{NS,g}= {\mathcal P}\left ( \chi_g \mu^{NS} + f_{1,g} \Theta^{NS}_{1\over 16}+ f_{7,g}\Theta^{NS}_{0} \right),
\ee
with coefficients of
\be
f_{r,g}(\t) = a'_r q^{-r^2/120} + \sum_{n=1}^\infty (\tr_{V_{r,n}^G}g )q^{n-r^2/120}~,~~a'_r= \begin{cases} -1& r=1 \\ 1 & r=7\end{cases}
\ee
given by characters of the bi-graded $G$-module
\be\label{def:VG}
V^G = \bigoplus_{r=1,7}\bigoplus_{n=1}^\infty V_{r,n}^G. 
\ee
By construction, this $G$-module can be seen as arising from the SCFT viewed as an extended $\mathcal N=1$ CFT. The first few coefficients of the $q$-series $f_{r,g}$ for all conjugacy classes in $G$ for $G=M_{24}, Co_2, Co_3$ are recorded in Appendix \ref{app:Coefficient Tables}. See also Appendix \ref{app:decomposition}  for the decomposition of the first few representations $V_{r,n}^G$ into irreducible $G$-representations.

We can now determine the modular properties of ${\underline f}_g :=  (f_{1,g},f_{7,g})$. 
 For $\chi_g\neq0$, the above analysis shows that ${\underline f}_g$  is a weakly holomorphic weight-${1/ 2}$ vector-valued mock modular form for $\Gamma_0(n_g)$, $n_g={\rm ord}(g)$,  with shadow $\chi_g\underline S(\t)$. In particular it has the multiplier $\bar \sigma : \Gamma_0(n_g) \to GL_2(\CC)$,  given by the inverse of the multiplier (\ref{multi_shadow}) restricted to the Hecke congruence subgroup $\Gamma_0(n_g)$. Recall that Hecke congruence subgroup is defined as 
  \[\Gamma_0(N)= \left\{ \begin{pmatrix} a & b\\ cN&d\end{pmatrix}\in SL_2(\ZZ) \Big\lvert a,b,c,d\in \ZZ \right\}.
  \]
  
For those conjugacy classes of $G\in \{M_{24}, Co_2, Co_3\}$ with  
  $\chi_g=0$, namely when the corresponding Frame shape (see Appendix \ref{sec:Frame Shapes})
  $$\Pi_g=\prod_{n\in{\cal I}_g} n^{k(n)}, ~~k(n)\neq 0$$ has $1\not\in {\cal I}_g$ and hence no (anti-)fixed point,  
  ${\underline f}_g$ is a vector-valued  weakly holomorphic weight-${1/ 2}$ mock modular form for $\Gamma_0(n_g)$.
  Its multiplier system is given by  
  \be
(c\t+d)^{-1/2}\, \rho_{\Pi_g}(\gamma) \bar \sigma(\g) \,\underline{f}(\tfrac{a\t+b}{c\t+d})  =  \underline{f}(\t) \quad,\quad {\rm for~all}~~\gamma=\bem a&b\\c&d \eem  \in \Gamma_0(n_g),
\ee where 
  $ \rho_{\Pi_g} : \Gamma_0(n_g) \to \CC^\ast$ is given by
  \be\label{def:multi_phase}
  \rho_{\Pi_g}(\gamma) = \begin{cases} e(-{cd \over n_g h_g}) , h_g={\rm min}\{n\vert n\in{\cal I}_g\}  &{\rm if}~~ k(n)>0~~ \forall~~ n\in {\cal I}_g \\ 
  1 &{\rm otherwise}.
  \end{cases}
  \ee
Namely, there is the above extra multiplying phase when the corresponding Frame shape is also a cycle shape. Note that for $G\cong M_{24}$, this is precisely the same multiplier system as the $M_{24}$ case of umbral moonshine. See \cite{Cheng:2012tq,UMNL,Cheng:2010pq,Gaberdiel:2010ca,Eguchi:2010fg}.

\section{Discussion}
\label{sec:Discussion}

In this paper and in previous work \cite{M5,Duncan}, we give examples of how sporadic groups can arise in a relatively simple theory. 
The relevant modular objects in the present work and in   \cite{M5}  are mock modular forms, and the appearance of these kinds of functions is directly related to the extended superconformal algebras that appear as infinite-dimensional symmetries of the theory. 
It is natural to wonder how generic this connection between mock modular forms and sporadic groups is, and how to construct these instances in general. 

It is also natural to wonder if there is any relation between the types of connections relating mock modular forms and sporadic groups discussed here, and those in umbral moonshine.
First, one might wonder if there is any relation between the $M_{24}$ module constructed in the present paper and the one underlying the $M_{24}$ umbral moonshine \cite{Eguchi2010,Gannon:2012ck}. 
As commented earlier, these two sets of functions have the same multiplier phases (\ref{def:multi_phase}) and hence correspond to the same cohomology class in the group cohomology $H^3(M_{24},\CC^\ast)$ (cf. \cite{Gaberdiel:2012gf,Gaberdiel:2013nya}). On the other hand, the resulting $M_{24}$ representations involved in the two modules seem to be distinct from each other. 
One also notes that the vector-valued function $(f_1,-f_7)$ has identical shadow (up to a scaling factor) and hence identical multiplier system  as the mock modular form underlying the umbral moonshine case corresponding to the Niemeier lattice $3E_8$.

Finally, another related construction \cite{DuncanMack-Crane2} based on the same free fermion orbifold theory leads to modules for subgroups of $\CoN$ fixing 4-planes. These are proposed to be related to symmetries of $K3$ sigma models. It would be interesting to explore what the connection, if any, between the sporadic group modules in the present work and $K3$ is. 

\bigskip
\centerline{\bf{Acknowledgements}}

We thank Nathan Benjamin, Natalie Paquette, and especially John Duncan for numerous conversations
about related subjects.
S.M.H. is supported by the Harvard University Golub Fellowship in the Physical Sciences.
S.K. and D.P.W. acknowledge the support of the NSF via grant PHY-0756174, the DoE
Office of Basic Energy Sciences contract DE-AC02-76SF00515, and the John Templeton
Foundation. 

\newpage

\appendix

\section{Jacobi Theta Functions}\label{sec:modforms}

We define the {\em Jacobi theta functions} $\th_i(\t,z)$ as follows for $q=e(\t)$ and $y=e(z)$:
\begin{align}	\th_1(\t,z)
	&= -i q^{1/8} y^{1/2} \prod_{n=1}^\infty (1-q^n) (1-y q^n) (1-y^{-1} q^{n-1})\,,\\
	\th_2(\t,z)
	&=  q^{1/8} y^{1/2} \prod_{n=1}^\infty (1-q^n) (1+y q^n) (1+y^{-1} q^{n-1})\,,\\
	\th_3(\t,z)
	&=  \prod_{n=1}^\infty (1-q^n) (1+y \,q^{n-1/2}) (1+y^{-1} q^{n-1/2})\,,\\
	\th_4(\t,z)
	&=  \prod_{n=1}^\infty (1-q^n) (1-y \,q^{n-1/2}) (1-y^{-1} q^{n-1/2})\,.
\end{align}

The weight four Eisenstein series $E_4$ can be written in terms of the Jacobi theta functions as
\be\label{def:E4}
E_4(\tau) = \frac12 \left( \th_2(\t,0)^8 + \th_3(\t,0)^8 + \th_4(\t,0)^8\right)\,.
\ee

\section{Characters}
\label{sec:characters}
\subsection*{B1. Frame Shapes and Spinor Representations}
\label{sec:Frame Shapes}

\begin{table}[h]
\begin{center}
\caption{Frame Shapes and Spinor Characters for $M_{24}$.}
\smallskip
\begin{footnotesize}
\begin{tabular}{c|ccccccccccc}\toprule
$[g]$       &   1A     &2A     &2B     &3A     &3B     &4A     &4B     &4C     &5A     &6A     &6B\\
\midrule
$\Pi_g$& $1^{24}$ & $1^82^8$ &$2^{12}$&$1^63^6$ &$3^8$&$2^44^4$&$1^42^24^4$ &$4^6$& $1^45^4$& $1^22^23^36^2$&$6^4$   \\
\midrule
Tr$_{\bf 4096}g$& 4096 & 0 &0& 64&16 & 0&0 & 0&16 & 0 &0
\\\midrule 
\midrule
$[g]$       &7AB    &8A     &10A    &11A    &12A    &12B    &14AB   &15AB   &21AB   &23AB\\
\midrule
$\Pi_g$&  $1^37^3$ & $1^22.4.8^2$  &$2^210^2$&$1^211^2$ & $2.4.6.12$&$12^2$& $1.2.7.14$& $1.3.5.15$&$3.21$&$1.23$     \\
\midrule
Tr$_{\bf 4096}g$& 8 & 0 & 0 & 4&0 & 0 & 0 & 4&2&2\\
\bottomrule
\end{tabular}
\end{footnotesize}
\end{center}
\end{table}

\begin{table}[h]
\begin{center}
\caption{Frame Shapes and Spinor Characters for $Co_{2}$.}
\smallskip
\scalebox{.85}{
\begin{footnotesize}
\begin{tabular}{c|cccccccccccc}\toprule
$[g]$          &1A     &2A     &2B   &2C  &3A     &3B     &4A     &4B  &4C & 4D & 4E & 4F   \\
\midrule
$\Pi_g$& $1^{24}$ &${2^{16}\over 1^8}$& $1^82^8$ &$2^{12}$&${3^9\over 1^3}$&$1^63^6$ &${1^84^8\over 2^8}$&${2^64^4\over 1^4}$&$1^42^24^4$ &${4^8\over 2^4}$&$1^42^24^4$ &$2^44^4$   \\
\midrule
Tr$_{\bf 4096}g$& 4096 & 0 &0& 0&-8 & 64&256 & 0&0 & 0 &0&0\\\midrule 
\midrule
$[g]$       & 4G    &5A  &5B   &6A     &6B &6C &6D &6E & 6F & 7A  &8A\\
\midrule
$\Pi_g$&  $4^6$ &${5^5\over 1}$&$1^45^4$&${1^53.6^4\over 2^4}$&${1.6^6\over 2^23^3}$&${1^42.6^5\over 3^4}$&${2^46^4\over 1^23^2}$&$1^22^23^26^2$&$2^36^3$&$1^37^3$&${2^48^4\over 4^4}$    \\
\midrule
Tr$_{\bf 4096}g$& 0 & -4 & 16 & 72&0 & 0 & 0 & 0&0&8&0\\\midrule 
\midrule
$[g]$        &8B & 8C & 8D & 8E & 8F & 9A   &10A  & 10B & 10C  &11A    &12A    \\
\midrule
$\Pi_g$&  ${2^34.8^2\over 1^2}$&${1^48^4\over 2^24^2}$&${8^4\over 4^2}$ & $1^22.4.8^2$  &$1^22.4.8^2$  &${1^39^3\over 3^2}$&${1^35.10^2\over 2^2}$&${1^22.10^3\over 5^2}$&$2^210^2$&$1^211^2$ & ${2.3^312^3\over 1.4.6^3}$     \\
\midrule
Tr$_{\bf 4096}g$& 0 & 32 & 0 & 0&0&4 & 20 & 0 & 0&4&4\\\midrule 
\midrule
$[g]$    &12B  &12C & 12D & 12E&12F & 12G & 12H&14A &14BC  &15A & 15BC \\
\midrule
$\Pi_g$& ${1.2^23.12^2\over 4^2}$&${1^23^24^212^2\over 2^26^2}$&${2^23^24.12\over 1^2}$&${1^312^3\over 2.3.4.6}$&${1^24.6^212\over 3^2}$&${1.2^23.12^2\over 4^2}$ & $2.4.6.12$&${2^214^2\over 1.7}$&$1.2.7.14$& $1.3.5.15$&${1^215^2\over 3.5}$     \\
\midrule
Tr$_{\bf 4096}g$& 0 & 16 & 0 & 12&0 & 0 & 0 & 0&0&4&2\\\midrule 
\midrule
$[g]$    & 16A & 16B & 18A &20A & 20B     &23AB & 24A & 24B & 28A & 30A & 30BC \\
\midrule
$\Pi_g$& ${1^216^2\over 2.8}$&${2^216^2\over 4.8}$&${1.2.18^2\over 6.9}$&${1.2.10.20\over 4.5}$&$4.20$&$1.23$ &${2.6.8.24\over 4.12}$&${1.4.6.24\over 3.8}$&${1.4.7.28\over 2.14}$&${2.3.5.30\over 1.15}$&${2.3.5.30\over 6.10}$   \\
\midrule
Tr$_{\bf 4096}g$& 8 & 0 & 0 & 0&0 & 2 & 0 & 0&4&0&2\\
\bottomrule
\end{tabular}
\end{footnotesize}}
\end{center}
\end{table}

\begin{table}[h]
\begin{center}
\caption{Frame Shapes and Spinor Characters for $Co_{3}$.}
\smallskip\scalebox{.85}{
\begin{footnotesize}
\begin{tabular}{c|ccccccccccccc}\toprule
$[g]$    &1A     &2A     &2B     &3A     &3B   &3C  &4A     &4B      &5A  &5B   &6A     &6B &6C   \\
\midrule

$\Pi_g$& $1^{24}$ & $1^82^8$ &$2^{12}$&${3^9\over 1^3}$&$1^63^6$ &$3^8$&${2^64^4\over 1^4}$&$1^42^24^4$ &${5^5\over 1}$& $1^45^4$& ${1^53.6^4\over 2^4}$&${3^36^3\over 1.2}$&$1^22^23^26^2$   \\
\midrule
Tr$_{\bf 4096}g$& 4096 & 0 &0& -8&64&16 & 0&0 & -4&16 & 72 &0&0
\\\midrule 
\midrule
$[g]$    &6D & 6E &7A  &8A  &8B & 8C   & 9A & 9B   &10A  & 10B   &11AB    &12A    &12B   \\
\midrule
$\Pi_g$&  $2^36^3$&$6^4$&$1^37^3$ & $1^22.4.8^2$  &${2^34.8^2\over 1^2}$&$1^22.4.8^2$&${9^3\over 3}$&${1^39^3\over 3^2}$&${1^35.10^2\over 2^2}$&$2^210^2$&$1^211^2$ & ${2^36.12^2\over 1.3.4^2}$&${1.2^23.12^2\over 4^2}$     \\
\midrule
Tr$_{\bf 4096}g$&0&0& 8 & 0 & 0 & 0&-2 & 4 & 20&0 & 4&0&0\\\midrule 
\midrule
$[g]$        &12C  &14A   &15A & 15B & 18A &20AB   &21A & 22AB   &23AB & 24A & 24B & 30A   \\
\midrule
$\Pi_g$&  ${1^24.6^212\over 3^2}$ &$1.2.7.14$&${1^215^2\over 3.5}$& $1.3.5.15$&${1^29.18\over 2.3}$&${1.2.10.20\over 4.5}$&$3.21$&$2.22$&$1.23$&${2.3.4.24\over 1.8}$&${1.4.6.24\over 3.8}$&${2.3.5.30\over 6.10}$     \\
\midrule
Tr$_{\bf 4096}g$& 0 & 0 & 2 & 4&6 & 0&2 & 0 & 2&0&0&2\\
\bottomrule
\end{tabular}
\end{footnotesize}}
\end{center}
\end{table}

\clearpage

\begin{sidewaystable}
\subsection*{B2. Irreducible Characters}\label{sec:chars:irr}
\begin{center}
\caption{Character table of $M_{24}$}\label{tab:chars:irr:2}
\smallskip
\begin{small}
\begin{tabular}{c@{ }|c|@{ }r@{ }r@{ }r@{ }r@{ }r@{ }r@{ }r@{ }r@{ }r@{ }r@{ }r@{ }r@{ }r@{ }r@{ }r@{ }r@{ }r@{ }r@{ }r@{ }r@{ }r@{ }r@{ }r@{ }r@{ }r@{ }r} \toprule
$[g]$	&{FS}&1A	&2A	&2B	&3A	&3B	&4A	&4B	&4C	&5A	&6A	&6B	&7A	&7B	&8A	&10A	&11A	&12A	&12B	&14A	&14B	&15A	&15B	&21A	&21B	&23A	&23B	\\
	\midrule
$[g^{2}]$	&&1A	&1A	&1A	&3A	&3B	&2A	&2A	&2B	&5A	&3A	&3B	&7A	&7B	&4B	&5A	&11A	&6A	&6B	&7A	&7B	&15A	&15B	&21A	&21B	&23A	&23B	\\
$[g^{3}]$	&&1A	&2A	&2B	&1A	&1A	&4A	&4B	&4C	&5A	&2A	&2B	&7B	&7A	&8A	&10A	&11A	&4A	&4C	&14B	&14A	&5A	&5A	&7B	&7A	&23A	&23B	\\
$[g^{5}]$	&&1A	&2A	&2B	&3A	&3B	&4A	&4B	&4C	&1A	&6A	&6B	&7B	&7A	&8A	&2B	&11A	&12A	&12B	&14B	&14A	&3A	&3A	&21B	&21A	&23B	&23A	\\
$[g^{7}]$	&&1A	&2A	&2B	&3A	&3B	&4A	&4B	&4C	&5A	&6A	&6B	&1A	&1A	&8A	&10A	&11A	&12A	&12B	&2A	&2A	&15B	&15A	&3B	&3B	&23B	&23A	\\
$[g^{11}]$	&&1A	&2A	&2B	&3A	&3B	&4A	&4B	&4C	&5A	&6A	&6B	&7A	&7B	&8A	&10A	&1A	&12A	&12B	&14A	&14B	&15B	&15A	&21A	&21B	&23B	&23A	\\
$[g^{23}]$	&&1A	&2A	&2B	&3A	&3B	&4A	&4B	&4C	&5A	&6A	&6B	&7A	&7B	&8A	&10A	&11A	&12A	&12B	&14A	&14B	&15A	&15B	&21A	&21B	&1A	&1A	\\
	\midrule
$\chi_{1}$	&$+$&$1$	&$1$	&$1$	&$1$	&$1$	&$1$	&$1$	&$1$	&$1$	&$1$	&$1$	&$1$	&$1$	&$1$	&$1$	&$1$	&$1$	&$1$	&$1$	&$1$	&$1$	&$1$	&$1$	&$1$	&$1$	&$1$	\\
\bottomrule
\end{tabular}
\end{small}
\end{center}
\end{sidewaystable}

\begin{sidewaystable}
\begin{center}
\caption{Character table of $Co_3$.}
\smallskip
\scalebox{.85}{
\begin{footnotesize}
\begin{tabular}{c|rrrrrrrrrrrrrrrrrrrrrrrrrrrrrrrrrrrrrrrrrrr}\toprule
$[g]$&1A&2A&2B&3A&3B&3C&4A&4B&5A&5B&6A&6B&6C&6D&6E&7A&8A&8B&8C&9A&9B\\
$[g^2]$&1A&1A&1A&3A&3B&3C&2A&2A&5A&5B&3A&3A&3B&3B&3C&7A&4B&4B&4B&9A&9B\\
$[g^3]$&1A&2A&2B&1A&1A&1A&4A&4B&5A&5B&2A&2A&2A&2B&2B&7A&8A&8B&8C&3A&3A\\
$[g^5]$&1A&2A&2B&3A&3B&3C&4A&4B&1A&1A&6A&6B&6C&6D&6E&7A&8A&8B&8C&9A&9B\\
$[g^7]$&1A&2A&2B&3A&3B&3C&4A&4B&5A&5B&6A&6B&6C&6D&6E&1A&8A&8B&8C&9A&9B\\
$[g^{11}]$&1A&2A&2B&3A&3B&3C&4A&4B&5A&5B&6A&6B&6C&6D&6E&7A&8A&8B&8C&9A&9B\\
$[g^{13}]$&1A&2A&2B&3A&3B&3C&4A&4B&5A&5B&6A&6B&6C&6D&6E&7A&8A&8B&8C&9A&9B\\
\midrule
${\chi}_{1}$&1&1&1&1&1&1&1&1&1&1&1&1&1&1&1&1&1&1&1&1&1\\
\bottomrule
\end{tabular}
\end{footnotesize}}
\end{center}
\end{sidewaystable}
\clearpage
\begin{sidewaystable}
\begin{center}
\caption{Character table of $Co_3$.}
\smallskip
\scalebox{.85}{
\begin{footnotesize}
\begin{tabular}{c|rrrrrrrrrrrrrrrrrrrrrrrrrrrrrrrrrrrrrrrrrrr}\toprule
$[g]$&10A&10B&11A&11B&12A&12B&12C&14A&15A&15B&18A&20A&20B&21A&22A&22B&23A&23B&24A&24B&30A\\
$[g^2]$&5A&5B&11B&11A&6A&6A&6C&7A&15A&15B&9A&10A&10A&21A&11B&11A&23A&23B&12B&12B&15A\\
$[g^3]$&10A&10B&11A&11B&4A&4B&4A&14A&5A&5B&6B&20A&20B&7A&22A&22B&23A&23B&8A&8B&10A\\
$[g^5]$&2A&2B&11A&11B&12A&12B&12C&14A&3A&3B&18A&4A&4A&21A&22A&22B&23B&23A&24A&24B&6A\\
$[g^7]$&10A&10B&11B&11A&12A&12B&12C&2A&15A&15B&18A&20A&20B&3C&22B&22A&23B&23A&24A&24B&30A\\
$[g^{11}]$&10A&10B&1A&1A&12A&12B&12C&14A&15A&15B&18A&20B&20A&21A&2B&2B&23B&23A&24A&24B&30A\\
$[g^{13}]$&10A&10B&11A&11B&12A&12B&12C&14A&15A&15B&18A&20A&20B&21A&22A&22B&1A&1A&24A&24B&30A\\
\midrule
${\chi}_{1}$&1&1&1&1&1&1&1&1&1&1&1&1&1&1&1&1&1&1&1&1&1\\
\bottomrule
\end{tabular}
\end{footnotesize}}
\end{center}
\end{sidewaystable}
\clearpage
\clearpage

\begin{table}
\begin{center}
\caption{Character table of $Co_2$.}
\scalebox{.75}{
\smallskip
\begin{footnotesize}
\begin{tabular}{c|rrrrrrrrrrrrrrrrrrrrrrrrrrrrrrrrrrrrrrrrrrrrrrrrrrrrrrrrrrrrr}\toprule
$[g]$&1A&2A&2B&2C&3A&3B&4A&4B&4C&4D&4E&4F&4G&5A&5B\\
$[g^2]$&1A&1A&1A&1A&3A&3B&2A&2B&2B&2A&2B&2B&2C&5A&5B\\
$[g^3]$&1A&2A&2B&2C&1A&1A&4A&4B&4C&4D&4E&4F&4G&5A&5B\\
$[g^5]$&1A&2A&2B&2C&3A&3B&4A&4B&4C&4D&4E&4F&4G&1A&1A\\
$[g^7]$&1A&2A&2B&2C&3A&3B&4A&4B&4C&4D&4E&4F&4G&5A&5B\\
$[g^{11}]$&1A&2A&2B&2C&3A&3B&4A&4B&4C&4D&4E&4F&4G&5A&5B\\
$[g^{13}]$&1A&2A&2B&2C&3A&3B&4A&4B&4C&4D&4E&4F&4G&5A&5B\\
\midrule
${\chi}_{1}$&1&1&1&1&1&1&1&1&1&1&1&1&1&1&1\\
\bottomrule
\end{tabular}
\end{footnotesize}}
\end{center}
\end{table}
\clearpage
\begin{table}
\begin{center}
\caption{Character table of $Co_2$.}
\smallskip
\scalebox{.75}{
\begin{footnotesize}
\begin{tabular}{c|rrrrrrrrrrrrrrrrrrrrrrrrrrrrrrrrrrrrrrrrrrrrrrrrrrrrrrrrrrrrr}\toprule
$[g]$&6A&6B&6C&6D&6E&6F&7A&8A&8B&8C&8D&8E&8F&9A&10A\\
$[g^2]$&3A&3A&3B&3B&3B&3B&7A&4A&4C&4D&4D&4C&4E&9A&5A\\
$[g^3]$&2B&2A&2A&2A&2B&2C&7A&8A&8B&8C&8D&8E&8F&3A&10A\\
$[g^5]$&6A&6B&6C&6D&6E&6F&7A&8A&8B&8C&8D&8E&8F&9A&2B\\
$[g^7]$&6A&6B&6C&6D&6E&6F&1A&8A&8B&8C&8D&8E&8F&9A&10A\\
$[g^{11}]$&6A&6B&6C&6D&6E&6F&7A&8A&8B&8C&8D&8E&8F&9A&10A\\
$[g^{13}]$&6A&6B&6C&6D&6E&6F&7A&8A&8B&8C&8D&8E&8F&9A&10A\\
\midrule
${\chi}_{1}$&1&1&1&1&1&1&1&1&1&1&1&1&1&1&1\\
\bottomrule
\end{tabular}
\end{footnotesize}}
\end{center}
\end{table}
\clearpage
\begin{table}
\begin{center}
\caption{Character table of $Co_2$.}
\smallskip
\scalebox{.75}{
\begin{footnotesize}
\begin{tabular}{c|rrrrrrrrrrrrrrrrrrrrrrrrrrrrrrrrrrrrrrrrrrrrrrrrrrrrrrrrrrrrr}\toprule
$[g]$&10B&10C&11A&12A&12B&12C&12D&12E&12F&12G&12H&14A&14B&14C&15A\\
$[g^2]$&5B&5B&11A&6B&6A&6D&6E&6B&6E&6A&6E&7A&7A&7A&15A\\
$[g^3]$&10B&10C&11A&4A&4C&4A&4C&4D&4B&4E&4F&14A&14C&14B&5B\\
$[g^5]$&2A&2C&11A&12A&12B&12C&12D&12E&12F&12G&12H&14A&14C&14B&3B\\
$[g^7]$&10B&10C&11A&12A&12B&12C&12D&12E&12F&12G&12H&2A&2B&2B&15A\\
$[g^{11}]$&10B&10C&1A&12A&12B&12C&12D&12E&12F&12G&12H&14A&14B&14C&15A\\
$[g^{13}]$&10B&10C&11A&12A&12B&12C&12D&12E&12F&12G&12H&14A&14B&a4C&15A\\
\midrule
${\chi}_{1}$&1&1&1&1&1&1&1&1&1&1&1&1&1&1&1\\
\bottomrule
\end{tabular}
\end{footnotesize}}
\end{center}
\end{table}
\clearpage
\begin{table}
\begin{center}
\caption{Character table of $Co_2$.}
\smallskip
\scalebox{.75}{
\begin{footnotesize}
\begin{tabular}{c|rrrrrrrrrrrrrrrrrrrrrrrrrrrrrrrrrrrrrrrrrrrrrrrrrrrrrrrrrrrrr}\toprule
$[g]$&15B&15C&16A&16B&18A&20A&20B&23A&23B&24A&24B&28A&30A&30B&30C\\
$[g^2]$&15B&15C&8D&8C&9A&10A&10C&23A&23B&12C&12B&14A&15A&15B&15C\\
$[g^3]$&5A&5A&16A&16B&6B&20A&20B&23A&23B&8A&8B&28A&10B&10A&10A\\
$[g^5]$&3A&3A&16A&16B&18A&4B&4G&23B&23A&24A&24B&28A&6C&6A&6A\\
$[g^7]$&15C&15B&16A&16B&18A&20A&20B&23B&23A&24A&24B&4A&30A&30C&30B\\
$[g^{11}]$&15C&15B&16A&16B&18A&20A&20B&23B&23A&24A&24B&28A&30A&30C&30B\\
$[g^{13}]$&15B&15C&16A&16B&18A&20A&20B&1A&1A&24A&24B&28A&30A&30B&30C\\
\midrule
${\chi}_{1}$&1&1&1&1&1&1&1&1&1&1&1&1&1&1&1\\
\bottomrule
\end{tabular}
\end{footnotesize}}
\end{center}
\end{table}
\clearpage

\begin{sidewaystable}
\section{Coefficient Tables}
\label{app:Coefficient Tables}
\begin{center}
\begin{small}

\caption{$M_{24}$ Twined series component $f_1(\t)$.}
\begin{tabular}{c@{ }|@{\;}r@{ }r@{ }r@{ }r@{ }r@{ }r@{ }r@{ }r@{ }r@{ }r@{ }r@{ }r@{ }r@{ }r@{ }r@{ }r@{ }r@{ }r@{ }r@{ }r@{ }r@{ }r@{ }r@{ }r@{ }r@{ }r}\toprule
$q^{n/120}$   &1A     &2A     &2B     &3A     &3B     &4A     &4B     &4C     &5A     &6A     &6B     &7AB    &8A     &10A    &11A    &12A    &12B    &14AB   &15AB   &21AB   &23AB   \\
\midrule
-1& -1 & -1 & -1 & -1 & -1 & -1 & -1 & -1 & -1 & -1 & -1 & -1 & -1 & -1 & -1 & -1 & -1 & -1 & -1 & -1 & -1 \\
119& 1771 & -21 & 11 & 16 & 7 & 3 & -5 & -1 & 1 & 0 & -1 & 0 & -1 & 1 & 0 & 0 & -1 & 0 & 1 & 0 & 0 \\
239& 35650 & 50 & -54 & 64 & -8 & 2 & -2 & 6 & 5 & -4 & 0 & -1 & 0 & 1 & -1 & 2 & 0 & 1 & -1 & -1 & 0 \\
359& 374141 & -147 & 165 & 119 & -13 & -11 & 9 & 5 & -4 & 3 & 3 & -2 & -1 & 0 & -2 & 1 & -1 & 0 & -1 & 1 & 0 \\
479& 2794156 & 412 & -396 & 277 & 43 & -4 & 8 & -16 & 11 & 1 & 3 & 1 & 2 & -1 & 2 & -1 & -1 & -1 & 2 & 1 & 1 \\
599& 16659038 & -946 & 902 & 668 & -49 & 22 & -22 & -10 & -2 & 8 & -1 & -3 & 0 & 2 & 0 & -2 & -1 & -1 & -2 & 0 & 0
   \\
719& 84505542 & 1974 & -2002 & 1032 & -36 & 14 & -14 & 30 & 2 & -12 & -4 & 2 & 0 & -2 & 0 & 2 & 0 & 0 & 2 & -1 & 0
   \\
839& 378649183 & -4017 & 4103 & 1975 & 196 & -41 & 43 & 15 & 18 & 3 & -4 & 3 & 1 & -2 & 0 & 1 & 0 & 1 & 0 & 0 & -1
   \\
959& 1536859149 & 7917 & -7875 & 3849 & -189 & -27 & 21 & -55 & -16 & -15 & 3 & 0 & -3 & 0 & 1 & -3 & -1 & 0 & -1 &
   0 & 0 \\
1079& 5749978586 & -14918 & 14762 & 5819 & -160 & 74 & -78 & -30 & 26 & 19 & 8 & -5 & -2 & 2 & 0 & -1 & 0 & -1 & -1 &
   1 & -1 \\
1199& 20088369809 & 26993 & -27071 & 9854 & 623 & 41 & -39 & 101 & -1 & -10 & 7 & -3 & 1 & -1 & 2 & 2 & -1 & 1 & -1 &
   0 & 0 \\
\bottomrule
\end{tabular}

\vspace{15pt}
\caption{$M_{24}$ Twined series component $f_7(\tau)$.}
\begin{tabular}{c@{ }|@{\;}r@{ }r@{ }r@{ }r@{ }r@{ }r@{ }r@{ }r@{ }r@{ }r@{ }r@{ }r@{ }r@{ }r@{ }r@{ }r@{ }r@{ }r@{ }r@{ }r@{ }r@{ }r@{ }r@{ }r@{ }r@{ }r}\toprule
$q^{n/120}$   &1A     &2A     &2B     &3A     &3B     &4A     &4B     &4C     &5A     &6A     &6B     &7AB    &8A     &10A    &11A    &12A    &12B    &14AB   &15AB   &21AB   &23AB   \\
\midrule
-49& 1 & 1 & 1 & 1 & 1 & 1 & 1 & 1 & 1 & 1 & 1 & 1 & 1 & 1 & 1 & 1 & 1 & 1 & 1 & 1 & 1 \\
71& 253 & 13 & -11 & 10 & 1 & -3 & 1 & 1 & 3 & -2 & 1 & 1 & -1 & -1 & 0 & 0 & 1 & -1 & 0 & 1 & 0 \\
191& 7359 & -49 & 55 & 33 & -15 & -1 & 3 & -5 & 4 & 5 & 1 & 2 & 1 & 0 & 0 & -1 & 1 & 0 & -2 & -1 & -1 \\
311& 95128 & 168 & -176 & 25 & 22 & 8 & -4 & -4 & 3 & -3 & -2 & -2 & 2 & -1 & 0 & -1 & 2 & 0 & 0 & 1 & 0 \\
431& 797226 & -454 & 450 & 177 & 21 & 2 & -2 & 10 & -4 & 5 & -3 & 3 & 0 & 0 & 1 & -1 & 1 & 1 & 2 & 0 & 0 \\
551& 5154853 & 1093 & -1067 & 301 & -98 & -11 & 13 & 5 & -2 & -11 & -2 & -3 & 1 & -2 & 0 & 1 & 2 & 1 & 1 & 0 & 1 \\
671& 27732365 & -2387 & 2397 & 386 & 104 & -11 & 5 & -15 & 10 & 10 & 0 & -4 & -3 & 2 & 1 & -2 & 0 & 0 & 1 & -1 & 0
   \\
791& 130136622 & 4942 & -4994 & 1149 & 84 & 22 & -26 & -6 & 17 & -11 & 4 & 0 & -2 & 1 & 0 & 1 & 0 & 0 & -1 & 0 & 0
   \\
911& 548143959 & -9849 & 9823 & 1929 & -399 & 15 & -13 & 31 & -1 & 21 & 1 & 6 & 1 & 3 & 0 & 3 & 1 & 0 & -1 & 0 & 1
   \\
1031& 2115036460 & 18780 & -18700 & 2431 & 391 & -44 & 40 & 20 & -20 & -21 & -1 & 5 & -2 & 0 & -2 & 1 & -1 & -1 & 1 &
   -1 & -1 \\
1151& 7584349199 & -34497 & 34551 & 5711 & 314 & -17 & 27 & -61 & -11 & 27 & -6 & -7 & 5 & 1 & -2 & 1 & 2 & -1 & 1 &
   -1 & 0 \\
   \bottomrule
\end{tabular}
\end{small}
\end{center}
\end{sidewaystable}

\begin{sidewaystable}
\begin{center}
\begin{small}
\caption{$Co_3$ Twined series component $f_1(\t)$.}
\begin{tabular}{c@{ }|@{\;}r@{ }r@{ }r@{ }r@{ }r@{ }r@{ }r@{ }r@{ }r@{ }r@{ }r@{ }r@{ }r@{ }r@{ }r@{ }r@{ }r@{ }r@{ }r@{ }r@{ }r@{ }r@{ }r@{ }r@{ }r@{ }rr@{ }r@{ }r@{ }r@{ }r@{ }r@{ }r@{ }r@{ }r@{ }r@{ }r@{ }r@{ }r@{ }r@{ }r@{ }r@{ }r@{ }r@{ }r@{ }r}\toprule
$q^{n/120}$  &1A     &2A     &2B     &3A     &3B   &3C  &4A     &4B      &5A  &5B   &6A     &6B &6C &6D & 6E &7A  \\
\midrule
-1&  -1 & -1 & -1 & -1 & -1 & -1 & -1 & -1 & -1 & -1 & -1 & -1 & -1 & -1 & -1 & -1 \\
119& 1771 & -21 & 11 & -11 & 16 & 7 & -5 & -5 & -4 & 1 & 21 & -3 & 0 & 2 & -1 & 0 \\
229& 35650 & 50 & -54 & 10 & 64 & -8 & -10 & -2 & 0 & 5 & 50 & -4 & -4 & 0 & 0 & -1 \\
359& 374141 & -147 & 165 & -25 & 119 & -13 & 1 & 9 & -9 & -4 & 135 & -3 & 3 & 3 & 3 & -2 \\
 479& 2794156 & 412 & -396 & -20 & 277 & 43 & 0 & 8 & 6 & 11 & 292 & -2 & 1 & -9 & 3 & 1 \\
599& 16659038 & -946 & 902 & 65 & 668 & -49 & -30 & -22 & -12 & -2 & 593 & -1 & 8 & 2 & -1 & -3 \\
719& 84505542 & 1974 & -2002 & -84 & 1032 & -36 & -22 & -14 & -8 & 2 & 1116 & -6 & -12 & -4 & -4 & 2 \\
839& 378649183 & -4017 & 4103 & -86 & 1975 & 196 & 35 & 43 & 8 & 18 & 2034 & -12 & 3 & 17 & -4 & 3 \\
959& 1536859149 & 7917 & -7875 & 276 & 3849 & -189 & 5 & 21 & -26 & -16 & 3564 & -12 & -15 & -9 & 3 & 0 \\
1079& 5749978586 & -14918 & 14762 & -301 & 5819 & -160 & -94 & -78 & 11 & 26 & 6091 & -5 & 19 & 11 & 8 & -5 \\
1199& 20088369809 & 26993 & -27071 & -226 & 9854 & 623 & -55 & -39 & -16 & -1 & 10094 & 2 & -10 & -26 & 7 & -3 \\
   \bottomrule
\end{tabular}
\vspace{25pt}

\begin{tabular}{c@{ }|@{\;}r@{ }r@{ }r@{ }r@{ }r@{ }r@{ }r@{ }r@{ }r@{ }r@{ }r@{ }r@{ }r@{ }r@{ }r@{ }r@{ }r@{ }r@{ }r@{ }r@{ }r@{ }r@{ }r@{ }r@{ }r@{ }r}\toprule
$q^{n/120}$    &8A  &8B & 8C & 9A & 9B   &10A  & 10B   &11AB    &12A    &12B  &12C  &14A   &15A & 15B & 18A &20AB   &21A & 22AB   &23AB & 24A & 24B & 30A   \\
\midrule
-1&  -1 & -1 & -1 & -1 & -1 & -1 & -1 & -1 & -1 & -1 & -1 & -1 & -1 & -1 & -1 & -1 & -1 & -1 & -1 & -1 & -1 & -1 \\
119&  -1 & -1 & -1 & -2 & -2 & 4 & 1 & 0 & 1 & 1 & -2 & 0 & -1 & 1 & 0 & 0 & 0 & 0 & 0 & -1 & -1 & 1 \\
239&  0 & -4 & 0 & 1 & 1 & 0 & 1 & -1 & -4 & -2 & 2 & 1 & 0 & -1 & -1 & 0 & -1 & 1 & 0 & 0 & 2 & 0 \\
359&  -1 & -5 & -1 & -4 & -1 & 3 & 0 & -2 & 1 & 3 & 1 & 0 & 0 & -1 & 0 & 1 & 1 & 0 & 0 & -1 & 1 & 0 \\
479& 2 & -2 & 2 & -2 & -2 & 2 & -1 & 2 & -6 & -4 & -3 & -1 & 0 & 2 & -2 & 0 & 1 & 0 & 1 & -4 & -2 & 2 \\
599& 0 & -4 & 0 & -4 & -1 & 4 & 2 & 0 & 3 & 5 & 0 & -1 & 0 & -2 & 2 & 0 & 0 & 0 & 0 & -3 & -1 & -2 \\
719& 0 & -4 & 0 & 0 & 3 & 4 & -2 & 0 & -10 & -8 & 2 & 0 & 1 & 2 & 0 & -2 & -1 & 0 & 0 & 0 & 2 & 1 \\
839& 1 & -3 & 1 & -2 & 1 & 8 & -2 & 0 & 8 & 10 & -1 & 1 & -1 & 0 & 0 & 0 & 0 & 0 & -1 & -2 & 0 & -1 \\
959& -3 & -11 & -3 & 0 & 3 & 2 & 0 & 1 & -16 & -12 & -1 & 0 & 1 & -1 & 0 & 0 & 0 & 1 & 0 & -6 & -2 & -1 \\
1079& -2 & -10 & -2 & -7 & -1 & 7 & 2 & 0 & 11 & 15 & -1 & -1 & -1 & -1 & 1 & 1 & 1 & 0 & -1 & -5 & -1 & 1 \\
1199& 1 & -7 & 1 & -7 & -4 & 8 & -1 & 2 & -22 & -18 & 2 & 1 & -1 & -1 & -1 & 0 & 0 & 0 & 0 & -2 & 2 & -1 \\ \bottomrule
\end{tabular}
\end{small}
\end{center}
\end{sidewaystable}

\begin{sidewaystable}
\begin{center}
\begin{small}

\caption{$Co_3$ Twined series component $f_7(\tau)$.}
\begin{tabular}{c@{ }|@{\;}r@{ }r@{ }r@{ }r@{ }r@{ }r@{ }r@{ }r@{ }r@{ }r@{ }r@{ }r@{ }r@{ }r@{ }r@{ }r@{ }r@{ }r@{ }r@{ }r@{ }r@{ }r@{ }r@{ }r@{ }r@{ }rr@{ }r@{ }r@{ }r@{ }r@{ }r@{ }r@{ }r@{ }r@{ }r@{ }r@{ }r@{ }r@{ }r@{ }r@{ }r@{ }r@{ }r@{ }r@{ }r}\toprule
$q^{n/120}$   &1A     &2A     &2B     &3A     &3B   &3C  &4A     &4B      &5A  &5B   &6A     &6B &6C &6D & 6E &7A  \\
\midrule
-49& 1 & 1 & 1 & 1 & 1 & 1 & 1 & 1 & 1 & 1 & 1 & 1 & 1 & 1 & 1 & 1 \\
71& 253 & 13 & -11 & 10 & 10 & 1 & 9 & 1 & 3 & 3 & 10 & 4 & -2 & -2 & 1 & 1 \\
191& 7359 & -49 & 55 & 15 & 33 & -15 & 11 & 3 & 9 & 4 & 23 & 5 & 5 & 1 & 1 & 2 \\
311& 95128 & 168 & -176 & -20 & 25 & 22 & 4 & -4 & 3 & 3 & 60 & 6 & -3 & -5 & -2 & -2 \\
431& 797226 & -454 & 450 & 51 & 177 & 21 & 6 & -2 & 1 & -4 & 131 & 5 & 5 & 9 & -3 & 3 \\
551& 5154853 & 1093 & -1067 & 40 & 301 & -98 & 29 & 13 & 3 & -2 & 280 & 4 & -11 & -5 & -2 & -3 \\
671& 27732365 & -2387 & 2397 & -145 & 386 & 104 & 21 & 5 & 15 & 10 & 535 & 7 & 10 & 12 & 0 & -4 \\
791& 130136622 & 4942 & -4994 & 186 & 1149 & 84 & -10 & -26 & 22 & 17 & 994 & 10 & -11 & -17 & 4 & 0 \\
911& 548143959 & -9849 & 9823 & 165 & 1929 & -399 & 11 & -13 & 9 & -1 & 1773 & 15 & 21 & 13 & 1 & 6 \\
1031& 2115036460 & 18780 & -18700 & -575 & 2431 & 391 & 64 & 40 & -15 & -20 & 3057 & 15 & -21 & -25 & -1 & 5 \\
1151& 7584349199 & -34497 & 34551 & 599 & 5711 & 314 & 51 & 27 & -1 & -11 & 5103 & 9 & 27 & 27 & -6 & -7 \\
   \bottomrule
\end{tabular}

\vspace{25pt}
\begin{tabular}{c@{ }|@{\;}r@{ }r@{ }r@{ }r@{ }r@{ }r@{ }r@{ }r@{ }r@{ }r@{ }r@{ }r@{ }r@{ }r@{ }r@{ }r@{ }r@{ }r@{ }r@{ }r@{ }r@{ }r@{ }r@{ }r@{ }r@{ }r}\toprule
$q^{n/120}$    &8A  &8B & 8C & 9A & 9B   &10A  & 10B   &11AB    &12A    &12B  &12C  &14A   &15A & 15B & 18A &20AB   &21A & 22AB   &23AB & 24A & 24B & 30A   \\
\midrule
-49& 1 & 1 & 1 & 1 & 1 & 1 & 1 & 1 & 1 & 1 & 1 & 1 & 1 & 1 & 1 & 1 & 1 & 1 & 1 & 1 & 1 & 1 \\
71& -1 & 3 & -1 & 1 & 1 & 3 & -1 & 0 & 0 & -2 & 0 & -1 & 0 & 0 & 1 & -1 & 1 & 0 & 0 & 2 & 0 & 0 \\
191& 1 & 5 & 1 & 3 & 0 & 1 & 0 & 0 & 5 & 3 & -1 & 0 & 0 & -2 & -1 & 1 & -1 & 0 & -1 & 1 & -1 & -2 \\
311& 2 & 6 & 2 & 4 & 1 & 3 & -1 & 0 & -2 & -4 & 1 & 0 & 0 & 0 & 0 & -1 & 1 & 0 & 0 & 2 & 0 & 0 \\
431& 0 & 4 & 0 & 3 & 0 & 1 & 0 & 1 & 9 & 7 & 3 & 1 & 1 & 2 & -1 & 1 & 0 & -1 & 0 & 3 & 1 & 1 \\
551& 1 & 9 & 1 & 1 & -2 & 3 & -2 & 0 & -4 & -8 & -1 & 1 & 0 & 1 & 1 & -1 & 0 & 0 & 1 & 4 & 0 & 0 \\
671& -3 & 5 & -3 & 5 & 2 & 3 & 2 & 1 & 15 & 11 & 0 & 0 & 0 & 1 & 1 & 1 & -1 & -1 & 0 & 3 & -1 & 0 \\
791& -2 & 6 & -2 & 3 & 0 & 2 & 1 & 0 & -10 & -14 & -1 & 0 & 1 & -1 & 1 & 0 & 0 & 0 & 0 & 4 & 0 & -1 \\
911& 1 & 13 & 1 & 6 & 0 & 1 & 3 & 0 & 23 & 17 & -1 & 0 & 0 & -1 & 0 & 1 & 0 & 0 & 1 & 7 & 1 & -2 \\
1031& -2 & 10 & -2 & 4 & -2 & 5 & 0 & -2 & -17 & -23 & 1 & -1 & 0 & 1 & 0 & -1 & -1 & 0 & -1 & 7 & 1 & 2 \\
1151&  5 & 17 & 5 & 8 & 2 & 3 & 1 & -2 & 33 & 27 & -3 & -1 & -1 & 1 & 0 & 1 & -1 & 0 & 0 & 5 & -1 & 3 \\\bottomrule
\end{tabular}
\end{small}
\end{center}
\end{sidewaystable}

\begin{sidewaystable}
\begin{center}
\begin{small}

\caption{$Co_2$ Twined series component $f_1(\t)$.}
\begin{tabular}{c@{ }|@{\;}r@{ }r@{ }r@{ }r@{ }r@{ }r@{ }r@{ }r@{ }r@{ }r@{ }r@{ }r@{ }r@{ }r@{ }r@{ }r@{ }r@{ }r@{ }r@{ }r@{ }r@{ }r@{ }r@{ }r@{ }r@{ }rr@{ }r@{ }r@{ }r@{ }r@{ }r@{ }r@{ }r@{ }r@{ }r@{ }r@{ }r@{ }r@{ }r@{ }r@{ }r@{ }r@{ }r@{ }r@{ }r}\toprule
$q^{n/120}$  &1A     &2A     &2B   &2C  &3A     &3B     &4A     &4B  &4C & 4D & 4E & 4F & 4G    &5A  &5B   &6A     &6B &6C &6D   \\
\midrule
-1&-1 & -1 & -1 & -1 & -1 & -1 & -1 & -1 & -1 & -1 & -1 & -1 & -1 & -1 & -1 & -1 & -1 & -1 & -1 \\
119& 1771 & -21 & -21 & 11 & -11 & 16 & 91 & -5 & -5 & -5 & -5 & 3 & -1 & -4 & 1 & 21 & -3 & -6 & 0 \\
239& 35650 & 34 & 50 & -54 & 10 & 64 & 482 & -10 & -2 & -6 & -2 & 2 & 6 & 0 & 5 & 50 & -2 & 4 & -8 \\
359& 374141 & -163 & -147 & 165 & -25 & 119 & 1997 & 1 & 9 & 5 & 9 & -11 & 5 & -9 & -4 & 135 & -1 & -1 & -1 \\
479& 2794156 & 396 & 412 & -396 & -20 & 277 & 6572 & 0 & 8 & 4 & 8 & -4 & -16 & 6 & 11 & 292 & 0 & 3 & -3 \\
599& 16659038 & -962 & -946 & 902 & 65 & 668 & 19118 & -30 & -22 & -26 & -22 & 22 & -10 & -12 & -2 & 593 & 1 & -2 &
   4 \\
719& 84505542 & 1958 & 1974 & -2002 & -84 & 1032 & 50806 & -22 & -14 & -18 & -14 & 14 & 30 & -8 & 2 & 1116 & -4 & 8
   & -16 \\
839& 378649183 & -4033 & -4017 & 4103 & -86 & 1975 & 125679 & 35 & 43 & 39 & 43 & -41 & 15 & 8 & 18 & 2034 & -10 &
   -19 & -1 \\
959& 1536859149 & 7885 & 7917 & -7875 & 276 & 3849 & 293053 & 5 & 21 & 13 & 21 & -27 & -55 & -26 & -16 & 3564 & -8 &
   7 & -23 \\
1079& 5749978586 & -14950 & -14918 & 14762 & -301 & 5819 & 651706 & -94 & -78 & -86 & -78 & 74 & -30 & 11 & 26 & 6091
   & -1 & -13 & 11 \\
1199& 20088369809 & 26961 & 26993 & -27071 & -226 & 9854 & 1392385 & -55 & -39 & -47 & -39 & 41 & 101 & -16 & -1 &
   10094 & 6 & 30 & -18 \\   \bottomrule
\end{tabular}
\vspace{25pt}

\begin{tabular}{c@{ }|@{\;}r@{ }r@{ }r@{ }r@{ }r@{ }r@{ }r@{ }r@{ }r@{ }r@{ }r@{ }r@{ }r@{ }r@{ }r@{ }r@{ }r@{ }r@{ }r@{ }r@{ }r@{ }r@{ }r@{ }r@{ }r@{ }r}\toprule
$q^{n/120}$  &6E & 6F & 7A  &8A  &8B & 8C & 8D & 8E & 8F & 9A   &10A  & 10B & 10C  &11A    &12A    &12B  &12C & 12D & 12E  \\
\midrule
-1& -1 & -1 & -1 & -1 & -1 & -1 & -1 & -1 & -1 & -1 & -1 & -1 & -1 & -1 & -1 & -1 & -1 & -1 & -1 \\
119& 0 & 2 & 0 & 3 & -1 & 7 & -1 & -1 & -1 & -2 & 4 & -1 & 1 & 0 & 1 & 1 & 4 & -2 & 1 \\
239& -4 & 0 & -1 & -6 & -4 & 10 & -2 & 0 & 0 & 1 & 0 & -1 & 1 & -1 & -4 & -2 & -4 & -2 & 0 \\
359& 3 & 3 & -2 & 13 & -5 & 17 & -3 & -1 & -1 & -1 & 3 & 2 & 0 & -2 & -1 & 3 & -1 & -3 & -1 \\
479& 1 & -9 & 1 & -20 & -2 & 28 & 0 & 2 & 2 & -2 & 2 & 1 & -1 & 2 & 2 & -4 & 5 & -7 & -2 \\
599& 8 & 2 & -3 & 30 & -4 & 50 & -2 & 0 & 0 & -1 & 4 & -2 & 2 & 0 & -7 & 5 & -4 & -4 & 1 \\
719& -12 & -4 & 2 & -50 & -4 & 66 & -2 & 0 & 0 & 3 & 4 & -2 & -2 & 0 & -8 & -8 & -8 & -2 & 0 \\
839& 3 & 17 & 3 & 71 & -3 & 99 & -1 & 1 & 1 & 1 & 8 & 2 & -2 & 0 & 12 & 10 & 15 & -5 & 0 \\
959& -15 & -9 & 0 & -107 & -11 & 137 & -7 & -3 & -3 & 3 & 2 & 0 & 0 & 1 & -14 & -12 & -11 & -9 & -2 \\
1079& 19 & 11 & -5 & 162 & -10 & 194 & -6 & -2 & -2 & -1 & 7 & 0 & 2 & 0 & -11 & 15 & -5 & -9 & 1 \\
1199& -10 & -26 & -3 & -223 & -7 & 269 & -3 & 1 & 1 & -4 & 8 & 1 & -1 & 2 & 22 & -18 & 22 & -6 & -2 \\ \bottomrule
\end{tabular}
\vspace{25pt}

\begin{tabular}{c@{ }|@{\;}r@{ }r@{ }r@{ }r@{ }r@{ }r@{ }r@{ }r@{ }r@{ }r@{ }r@{ }r@{ }r@{ }r@{ }r@{ }r@{ }r@{ }r@{ }r@{ }r@{ }r@{ }r@{ }r@{ }r@{ }r@{ }r}\toprule
$q^{n/120}$ &12F & 12G & 12H&14A &14BC  &15A & 15BC & 16A & 16B & 18A &20A & 20B     &23AB & 24A & 24B & 28A & 30A & 30BC  \\
\midrule
-1& -1 & -1 & -1 & -1 & -1 & -1 & -1 & -1 & -1 & -1 & -1 & -1 & -1 & -1 & -1 & -1 & -1 & -1 \\
119& -2 & 1 & 0 & 0 & 0 & 1 & -1 & 1 & 1 & 0 & 0 & -1 & 0 & 0 & -1 & 0 & -1 & 1 \\
239& 2 & -2 & 2 & -1 & 1 & -1 & 0 & -2 & 0 & 1 & 0 & 1 & 0 & 0 & 2 & -1 & -1 & 0 \\
359& 1 & 3 & 1 & -2 & 0 & -1 & 0 & -1 & 1 & -1 & 1 & 0 & 0 & 1 & 1 & 2 & -1 & 0 \\
479& -3 & -4 & -1 & -3 & -1 & 2 & 0 & 0 & -2 & 0 & 0 & -1 & 1 & 1 & -2 & -1 & -2 & 2 \\
599& 0 & 5 & -2 & -3 & -1 & -2 & 0 & 0 & 2 & 1 & 0 & 0 & 0 & 0 & -1 & 1 & -2 & -2 \\
719& 2 & -8 & 2 & -2 & 0 & 2 & 1 & 0 & -2 & -1 & -2 & 0 & 0 & -2 & 2 & 0 & -2 & 1 \\
 839&-1 & 10 & 1 & -1 & 1 & 0 & -1 & 1 & 3 & -1 & 0 & 0 & -1 & -1 & 0 & 1 & -4 & -1 \\
959& -1 & -12 & -3 & -4 & 0 & -1 & 1 & -1 & -1 & 1 & 0 & 0 & 0 & 1 & -2 & -2 & -3 & -1 \\
1079& -1 & 15 & -1 & -5 & -1 & -1 & -1 & 2 & 2 & -1 & 1 & 0 & -1 & 3 & -1 & -1 & -3 & 1 \\
1199& 2 & -18 & 2 & -3 & 1 & -1 & -1 & -1 & -5 & 0 & 0 & 1 & 0 & 2 & 2 & 1 & -5 & -1 \\ \bottomrule
\end{tabular}

\end{small}
\end{center}
\end{sidewaystable}

\begin{sidewaystable}
\begin{center}
\begin{small}

\caption{$Co_2$ Twined series component $f_7(\t)$.}
\begin{tabular}{c@{ }|@{\;}r@{ }r@{ }r@{ }r@{ }r@{ }r@{ }r@{ }r@{ }r@{ }r@{ }r@{ }r@{ }r@{ }r@{ }r@{ }r@{ }r@{ }r@{ }r@{ }r@{ }r@{ }r@{ }r@{ }r@{ }r@{ }rr@{ }r@{ }r@{ }r@{ }r@{ }r@{ }r@{ }r@{ }r@{ }r@{ }r@{ }r@{ }r@{ }r@{ }r@{ }r@{ }r@{ }r@{ }r@{ }r}\toprule
$q^{n/120}$   &1A     &2A     &2B   &2C  &3A     &3B     &4A     &4B  &4C & 4D & 4E & 4F & 4G    &5A  &5B   &6A     &6B &6C &6D   \\
\midrule
-49& 1 & 1 & 1 & 1 & 1 & 1 & 1 & 1 & 1 & 1 & 1 & 1 & 1 & 1 & 1 & 1 & 1 & 1 & 1 \\
71& 253 & 29 & 13 & -11 & 10 & 10 & 29 & 9 & 1 & 5 & 1 & -3 & 1 & 3 & 3 & 10 & 2 & 2 & 2 \\
191& 7359 & -33 & -49 & 55 & 15 & 33 & 159 & 11 & 3 & 7 & 3 & -1 & -5 & 9 & 4 & 23 & 3 & -3 & 9 \\
311& 95128 & 184 & 168 & -176 & -20 & 25 & 728 & 4 & -4 & 0 & -4 & 8 & -4 & 3 & 3 & 60 & 4 & 7 & 1 \\
431& 797226 & -438 & -454 & 450 & 51 & 177 & 2554 & 6 & -2 & 2 & -2 & 2 & 10 & 1 & -4 & 131 & 3 & -3 & 9 \\
551& 5154853 & 1125 & 1093 & -1067 & 40 & 301 & 7813 & 29 & 13 & 21 & 13 & -11 & 5 & 3 & -2 & 280 & 0 & 3 & -3 \\
671& 27732365 & -2355 & -2387 & 2397 & -145 & 386 & 21469 & 21 & 5 & 13 & 5 & -11 & -15 & 15 & 10 & 535 & 3 & -12 &
   18 \\
791& 130136622 & 4974 & 4942 & -4994 & 186 & 1149 & 54462 & -10 & -26 & -18 & -26 & 22 & -6 & 22 & 17 & 994 & 6 & 15
   & -3 \\
911& 548143959 & -9801 & -9849 & 9823 & 165 & 1929 & 129831 & 11 & -13 & -1 & -13 & 15 & 31 & 9 & -1 & 1773 & 9 &
   -15 & 33 \\
1031& 2115036460 & 18828 & 18780 & -18700 & -575 & 2431 & 294172 & 64 & 40 & 52 & 40 & -44 & 20 & -15 & -20 & 3057 &
   9 & 27 & -9 \\
1151& 7584349199 & -34449 & -34497 & 34551 & 599 & 5711 & 638063 & 51 & 27 & 39 & 27 & -17 & -61 & -1 & -11 & 5103 &
   3 & -33 & 39 \\
   \bottomrule
\end{tabular}
\vspace{25pt}

\begin{tabular}{c@{ }|@{\;}r@{ }r@{ }r@{ }r@{ }r@{ }r@{ }r@{ }r@{ }r@{ }r@{ }r@{ }r@{ }r@{ }r@{ }r@{ }r@{ }r@{ }r@{ }r@{ }r@{ }r@{ }r@{ }r@{ }r@{ }r@{ }r}\toprule
$q^{n/120}$  &6E & 6F & 7A  &8A  &8B & 8C & 8D & 8E & 8F & 9A   &10A  & 10B & 10C  &11A    &12A    &12B  &12C & 12D & 12E  \\
\midrule
-49&1 & 1 & 1 & 1 & 1 & 1 & 1 & 1 & 1 & 1 & 1 & 1 & 1 & 1 & 1 & 1 & 1 & 1 & 1 \\
71& -2 & -2 & 1 & -3 & 3 & 5 & 1 & -1 & -1 & 1 & 3 & -1 & -1 & 0 & 2 & -2 & 2 & 4 & 2 \\
191& 5 & 1 & 2 & 7 & 5 & 7 & 3 & 1 & 1 & 0 & 1 & 2 & 0 & 0 & -3 & 3 & -3 & 3 & 1 \\
311& -3 & -5 & -2 & -16 & 6 & 8 & 4 & 2 & 2 & 1 & 3 & -1 & -1 & 0 & 8 & -4 & 5 & 5 & 0 \\
431& 5 & 9 & 3 & 26 & 4 & 14 & 2 & 0 & 0 & 0 & 1 & 2 & 0 & 1 & 7 & 7 & 1 & 7 & -1 \\
551& -11 & -5 & -3 & -43 & 9 & 29 & 5 & 1 & 1 & -2 & 3 & 0 & -2 & 0 & -8 & -8 & -11 & 7 & 0 \\
671& 10 & 12 & -4 & 69 & 5 & 33 & 1 & -3 & -3 & 2 & 3 & 0 & 2 & 1 & 13 & 11 & 10 & 8 & 1 \\
791& -11 & -17 & 0 & -98 & 6 & 50 & 2 & -2 & -2 & 0 & 2 & -1 & 1 & 0 & 12 & -14 & 9 & 7 & 0 \\
911& 21 & 13 & 6 & 151 & 13 & 75 & 7 & 1 & 1 & 0 & 1 & -1 & 3 & 0 & -21 & 17 & -27 & 11 & -1 \\
1031& -21 & -25 & 5 & -220 & 10 & 104 & 4 & -2 & -2 & -2 & 5 & -2 & 0 & -2 & 25 & -23 & 19 & 13 & 1 \\
1151& 27 & 27 & -7 & 311 & 17 & 151 & 11 & 5 & 5 & 2 & 3 & 1 & 1 & -2 & 17 & 27 & 11 & 9 & -3 \\ \bottomrule
\end{tabular}
\vspace{25pt}

\begin{tabular}{c@{ }|@{\;}r@{ }r@{ }r@{ }r@{ }r@{ }r@{ }r@{ }r@{ }r@{ }r@{ }r@{ }r@{ }r@{ }r@{ }r@{ }r@{ }r@{ }r@{ }r@{ }r@{ }r@{ }r@{ }r@{ }r@{ }r@{ }r}\toprule
$q^{n/120}$ &12F & 12G & 12H&14A &14BC  &15A & 15BC & 16A & 16B & 18A &20A & 20B     &23AB & 24A & 24B & 28A & 30A & 30BC  \\
\midrule
-49& 1 & 1 & 1 & 1 & 1 & 1 & 1 & 1 & 1 & 1 & 1 & 1 & 1 & 1 & 1 & 1 & 1 & 1 \\
71& 0 & -2 & 0 & 1 & -1 & 0 & 0 & 1 & -1 & -1 & -1 & 1 & 0 & 0 & 0 & 1 & 2 & 0 \\
191& -1 & 3 & -1 & 2 & 0 & -2 & 0 & -1 & 1 & 0 & 1 & 0 & -1 & 1 & -1 & -2 & 2 & -2 \\
311& 1 & -4 & -1 & 2 & 0 & 0 & 0 & 0 & -2 & 1 & -1 & 1 & 0 & -1 & 0 & 0 & 2 & 0 \\
431& 3 & 7 & -1 & 3 & 1 & 2 & 1 & 0 & 2 & 0 & 1 & 0 & 0 & -1 & 1 & -1 & 2 & 1 \\
551&  -1 & -8 & 1 & 5 & 1 & 1 & 0 & 1 & -3 & 0 & -1 & 0 & 1 & -1 & 0 & 1 & 3 & 0 \\
671& 0 & 11 & -2 & 4 & 0 & 1 & 0 & -1 & 3 & 0 & 1 & 0 & 0 & 0 & -1 & 0 & 3 & 0 \\
791& -1 & -14 & 1 & 4 & 0 & -1 & 1 & 0 & -4 & 0 & 0 & -1 & 0 & 1 & 0 & 2 & 5 & -1 \\
911& -1 & 17 & 3 & 6 & 0 & -1 & 0 & 1 & 3 & 0 & 1 & 1 & 1 & 1 & 1 & 2 & 5 & -2 \\
1031& 1 & -23 & 1 & 5 & -1 & 1 & 0 & 2 & -4 & 0 & -1 & 0 & -1 & -1 & 1 & -3 & 7 & 2 \\
1151& -3 & 27 & 1 & 5 & -1 & 1 & -1 & -1 & 5 & 0 & 1 & -1 & 0 & -1 & -1 & -1 & 7 & 3 \\
 \bottomrule
\end{tabular}

\end{small}
\end{center}
\end{sidewaystable}

\begin{sidewaystable}
\section{Decomposition Tables}
\label{app:decomposition}
\begin{center}
\begin{small}
\caption{$M_{24}$ decompositions for $f_1(\tau)$.}
\begin{tabular}{c@{ }|@{\;}r@{ }r@{ }r@{ }r@{ }r@{ }r@{ }r@{ }r@{ }r@{ }r@{ }r@{ }r@{ }r@{ }r@{ }r@{ }r@{ }r@{ }r@{ }r@{ }r@{ }r@{ }r@{ }r@{ }r@{ }r@{ }r@{ }r@{ }r@{ }r@{ }r@{ }r}\toprule
$q^{n/120}$   &$\chi_1$   &$\chi_2$   &$\chi_3$   &$\chi_4$   &$\chi_5$   &$\chi_6$   &$\chi_7$   &$\chi_8$   &$\chi_9$   &$\chi_{10}$   &$\chi_{11}$   &$\chi_{12}$   &$\chi_{13}$     \\
\midrule
 -1&-1 & 0 & 0 & 0 & 0 & 0 & 0 & 0 & 0 & 0 & 0 & 0 & 0 \\
119& 0 & 0 & 0 & 0 & 0 & 0 & 0 & 0 & 0 & 0 & 0 & 0 & 0 \\
239& 0 & 0 & 0 & 0 & 0 & 0 & 1 & 1 & 0 & 0 & 0 & 0 & 0 \\
359& 0 & 0 & 0 & 0 & 0 & 0 & 2 & 1 & 2 & 2 & 2 & 1 & 1 \\
479& 1 & 3 & 0 & 0 & 2 & 2 & 5 & 7 & 7 & 9 & 9 & 12 & 12 \\
599& 0 & 4 & 3 & 3 & 13 & 13 & 23 & 20 & 35 & 59 & 59 & 67 & 67 \\
719& 1 & 13 & 14 & 14 & 80 & 80 & 94 & 102 & 175 & 266 & 266 & 342 & 342 \\
839& 5 & 45 & 75 & 75 & 346 & 346 & 409 & 403 & 754 & 1204 & 1204 & 1530 & 1530 \\
959& 7 & 166 & 275 & 275 & 1449 & 1449 & 1611 & 1640 & 3061 & 4839 & 4839 & 6218 & 6218 \\
1079& 30 & 560 & 1066 & 1066 & 5388 & 5388 & 5971 & 5964 & 11356 & 18146 & 18146 & 23244 & 23244 \\
1199& 91 & 1941 & 3675 & 3675 & 18970 & 18970 & 20751 & 20908 & 39721 & 63163 & 63163 & 81246 & 81246 \\
\bottomrule
\end{tabular}

\vspace{25pt}
\begin{tabular}{c@{ }|@{\;}r@{ }r@{ }r@{ }r@{ }r@{ }r@{ }r@{ }r@{ }r@{ }r@{ }r@{ }r@{ }r@{ }r@{ }r@{ }r@{ }r@{ }r@{ }r@{ }r@{ }r@{ }r@{ }r@{ }r@{ }r@{ }r@{ }r@{ }r@{ }r@{ }r@{ }r}\toprule
$q^{n/120}$  &$\chi_{14}$   &$\chi_{15}$   &$\chi_{16}$   &$\chi_{17}$   &$\chi_{18}$   &$\chi_{19}$   &$\chi_{20}$   &$\chi_{21}$   &$\chi_{22}$   &$\chi_{23}$   &$\chi_{24}$   &$\chi_{25}$   &$\chi_{26}$   \\
\midrule
-1& 0 & 0 & 0 & 0 & 0 & 0 & 0 & 0 & 0 & 0 & 0 & 0 & 0 \\
119& 0 & 0 & 0 & 0 & 1 & 0 & 0 & 0 & 0 & 0 & 0 & 0 & 0 \\
239& 0 & 0 & 0 & 0 & 1 & 0 & 0 & 0 & 2 & 0 & 1 & 0 & 2 \\
359& 2 & 2 & 2 & 2 & 4 & 3 & 3 & 5 & 6 & 6 & 11 & 9 & 15 \\
479& 11 & 12 & 12 & 19 & 24 & 22 & 28 & 37 & 43 & 58 & 63 & 62 & 120 \\
599& 73 & 71 & 71 & 85 & 133 & 139 & 151 & 226 & 243 & 352 & 388 & 393 & 703 \\
719& 351 & 356 & 356 & 447 & 622 & 692 & 791 & 1144 & 1232 & 1821 & 1910 & 1980 & 3599 \\
839& 1616 & 1601 & 1601 & 1952 & 2779 & 3142 & 3509 & 5119 & 5447 & 8185 & 8615 & 8974 & 16057 \\
959& 6467 & 6497 & 6497 & 7991 & 11152 & 12678 & 14322 & 20795 & 22160 & 33307 & 34789 & 36306 & 65290 \\
1079& 24356 & 24314 & 24314 & 29672 & 41718 & 47569 & 53424 & 77784 & 82682 & 124690 & 130340 & 136163 & 244068 \\
1199& 84847 & 84911 & 84911 & 103962 & 145405 & 166002 & 186931 & 271754 & 288988 & 435836 & 454826 & 475331 & 853073
   \\
\bottomrule
\end{tabular}
\end{small}
\end{center}

\end{sidewaystable}

\begin{sidewaystable}
\begin{center}
\begin{small}

\caption{$M_{24}$ decompositions for $f_7(\tau)$.}
\begin{tabular}{c@{ }|@{\;}r@{ }r@{ }r@{ }r@{ }r@{ }r@{ }r@{ }r@{ }r@{ }r@{ }r@{ }r@{ }r@{ }r@{ }r@{ }r@{ }r@{ }r@{ }r@{ }r@{ }r@{ }r@{ }r@{ }r@{ }r@{ }r@{ }r@{ }r@{ }r@{ }r@{ }r}\toprule
$q^{n/120}$   &$\chi_1$   &$\chi_2$   &$\chi_3$   &$\chi_4$   &$\chi_5$   &$\chi_6$   &$\chi_7$   &$\chi_8$   &$\chi_9$   &$\chi_{10}$   &$\chi_{11}$   &$\chi_{12}$   &$\chi_{13}$    \\
\midrule
-49& 1 & 0 & 0 & 0 & 0 & 0 & 0 & 0 & 0 & 0 & 0 & 0 & 0 \\
71& 0 & 0 & 0 & 0 & 0 & 0 & 0 & 1 & 0 & 0 & 0 & 0 & 0 \\
191& 0 & 1 & 0 & 0 & 0 & 0 & 1 & 0 & 0 & 1 & 1 & 0 & 0 \\
311& 0 & 0 & 0 & 0 & 0 & 0 & 0 & 1 & 0 & 0 & 0 & 1 & 1 \\
431& 1 & 1 & 1 & 1 & 0 & 0 & 2 & 1 & 3 & 4 & 4 & 3 & 3 \\
551& 0 & 2 & 0 & 0 & 5 & 5 & 7 & 11 & 13 & 16 & 16 & 21 & 21 \\
671& 1 & 4 & 8 & 8 & 22 & 22 & 33 & 27 & 54 & 93 & 93 & 112 & 112 \\
791& 1 & 19 & 20 & 20 & 127 & 127 & 142 & 157 & 267 & 404 & 404 & 529 & 529 \\
911& 6 & 57 & 106 & 106 & 499 & 499 & 584 & 562 & 1082 & 1758 & 1758 & 2209 & 2209 \\
1031& 9 & 217 & 377 & 377 & 2016 & 2016 & 2192 & 2249 & 4210 & 6619 & 6619 & 8563 & 8563 \\
1151& 40 & 724 & 1423 & 1423 & 7087 & 7087 & 7865 & 7817 & 14953 & 23945 & 23945 & 30657 & 30657 \\
\bottomrule
\end{tabular}

\vspace{25pt}
\begin{tabular}{c@{ }|@{\;}r@{ }r@{ }r@{ }r@{ }r@{ }r@{ }r@{ }r@{ }r@{ }r@{ }r@{ }r@{ }r@{ }r@{ }r@{ }r@{ }r@{ }r@{ }r@{ }r@{ }r@{ }r@{ }r@{ }r@{ }r@{ }r@{ }r@{ }r@{ }r@{ }r@{ }r}\toprule
$q^{n/120}$   &$\chi_{14}$   &$\chi_{15}$   &$\chi_{16}$   &$\chi_{17}$   &$\chi_{18}$   &$\chi_{19}$   &$\chi_{20}$   &$\chi_{21}$   &$\chi_{22}$   &$\chi_{23}$   &$\chi_{24}$   &$\chi_{25}$   &$\chi_{26}$   \\
\midrule
 -49&0 & 0 & 0 & 0 & 0 & 0 & 0 & 0 & 0 & 0 & 0 & 0 & 0 \\
 71&0 & 0 & 0 & 0 & 0 & 0 & 0 & 0 & 0 & 0 & 0 & 0 & 0 \\
 191&0 & 0 & 0 & 0 & 0 & 0 & 0 & 0 & 0 & 0 & 1 & 0 & 0 \\
311& 0 & 0 & 0 & 2 & 1 & 1 & 1 & 1 & 2 & 2 & 1 & 1 & 5 \\
431& 5 & 3 & 3 & 3 & 10 & 8 & 6 & 11 & 11 & 14 & 22 & 20 & 32 \\
 551&17 & 22 & 22 & 31 & 37 & 38 & 50 & 71 & 83 & 108 & 113 & 114 & 224 \\
671& 126 & 118 & 118 & 138 & 214 & 236 & 251 & 373 & 393 & 595 & 645 & 668 & 1165 \\
791& 535 & 548 & 548 & 699 & 949 & 1063 & 1228 & 1759 & 1896 & 2815 & 2927 & 3041 & 5550 \\
911& 2346 & 2322 & 2322 & 2792 & 4012 & 4550 & 5059 & 7419 & 7874 & 11858 & 12486 & 13018 & 23227 \\
1031& 8884 & 8933 & 8933 & 11024 & 15294 & 17438 & 19742 & 28612 & 30483 & 45883 & 47808 & 49942 & 89894 \\
1151& 32179 & 32073 & 32073 & 39078 & 55035 & 62794 & 70425 & 102594 & 108989 & 164475 & 171994 & 179713 & 321855 \\
\bottomrule
\end{tabular}
\end{small}
\end{center}

\end{sidewaystable}

\begin{sidewaystable}
\begin{center}
\begin{small}
\caption{$Co_3$ decompositions for $f_1(\tau)$.}
\begin{tabular}{c@{ }|@{\;}r@{ }r@{ }r@{ }r@{ }r@{ }r@{ }r@{ }r@{ }r@{ }r@{ }r@{ }r@{ }r@{ }r@{ }r@{ }r@{ }r@{ }r@{ }r@{ }r@{ }r@{ }r@{ }r@{ }r@{ }r@{ }r@{ }r@{ }r@{ }r@{ }r@{ }r}\toprule
$q^{n/120}$   &$\chi_1$   &$\chi_2$   &$\chi_3$   &$\chi_4$   &$\chi_5$   &$\chi_6$   &$\chi_7$   &$\chi_8$   &$\chi_9$   &$\chi_{10}$   &$\chi_{11}$   &$\chi_{12}$   &$\chi_{13}$  &$\chi_{14}$   &$\chi_{15}$   &$\chi_{16}$   &$\chi_{17}$   &$\chi_{18}$   &$\chi_{19}$   &$\chi_{20}$   &$\chi_{21}$    \\
\midrule
-1& -1 & 0 & 0 & 0 & 0 & 0 & 0 & 0 & 0 & 0 & 0 & 0 & 0 & 0 & 0 & 0 & 0 & 0 & 0 & 0 & 0 \\
119& 0 & 0 & 0 & 0 & 0 & 0 & 0 & 1 & 0 & 0 & 0 & 0 & 0 & 0 & 0 & 0 & 0 & 0 & 0 & 0 & 0 \\
239& 0 & 0 & 0 & 0 & 0 & 0 & 0 & 0 & 0 & 0 & 0 & 1 & 0 & 0 & 0 & 0 & 0 & 0 & 0 & 0 & 0 \\
359& 0 & 0 & 0 & 0 & 0 & 0 & 0 & 1 & 0 & 0 & 0 & 1 & 0 & 1 & 0 & 0 & 0 & 0 & 0 & 1 & 0 \\
479& 0 & 1 & 1 & 1 & 0 & 0 & 0 & 2 & 1 & 0 & 0 & 1 & 1 & 1 & 0 & 0 & 0 & 0 & 0 & 0 & 1 \\
599& 0 & 1 & 1 & 0 & 1 & 0 & 0 & 4 & 0 & 0 & 0 & 3 & 1 & 4 & 0 & 0 & 0 & 0 & 0 & 3 & 2 \\
719& 0 & 2 & 4 & 1 & 1 & 0 & 0 & 6 & 1 & 0 & 0 & 6 & 3 & 5 & 0 & 2 & 2 & 3 & 3 & 5 & 7 \\
839& 1 & 3 & 5 & 2 & 5 & 0 & 0 & 14 & 3 & 1 & 1 & 12 & 8 & 16 & 6 & 6 & 6 & 15 & 15 & 26 & 23 \\
959& 0 & 5 & 11 & 4 & 5 & 2 & 2 & 24 & 9 & 9 & 9 & 30 & 25 & 37 & 23 & 31 & 31 & 61 & 61 & 75 & 91 \\
1079& 2 & 7 & 17 & 7 & 14 & 11 & 11 & 56 & 26 & 38 & 38 & 74 & 74 & 117 & 100 & 106 & 106 & 234 & 234 & 290 & 311 \\
1199& 2 & 13 & 40 & 20 & 24 & 32 & 32 & 125 & 91 & 135 & 135 & 208 & 245 & 326 & 343 & 398 & 398 & 831 & 831 & 942 &
   1084 \\
\bottomrule
\end{tabular}

\vspace{25pt}
\begin{tabular}{c@{ }|@{\;}r@{ }r@{ }r@{ }r@{ }r@{ }r@{ }r@{ }r@{ }r@{ }r@{ }r@{ }r@{ }r@{ }r@{ }r@{ }r@{ }r@{ }r@{ }r@{ }r@{ }r@{ }r@{ }r@{ }r@{ }r@{ }r@{ }r@{ }r@{ }r@{ }r@{ }r}\toprule
$q^{n/120}$     &$\chi_{22}$   &$\chi_{23}$   &$\chi_{24}$   &$\chi_{25}$   &$\chi_{26}$ &$\chi_{27}$   &$\chi_{28}$   &$\chi_{29}$   &$\chi_{30}$   &$\chi_{31}$  &$\chi_{32}$   &$\chi_{33}$  &$\chi_{34}$   &$\chi_{35}$   &$\chi_{36}$   &$\chi_{37}$   &$\chi_{38}$   &$\chi_{39}$   &$\chi_{40}$   &$\chi_{41}$ &$\chi_{42}$  \\
\midrule
-1& 0 & 0 & 0 & 0 & 0 & 0 & 0 & 0 & 0 & 0 & 0 & 0 & 0 & 0 & 0 & 0 & 0 & 0 & 0 & 0 & 0 \\
119& 0 & 0 & 0 & 0 & 0 & 0 & 0 & 0 & 0 & 0 & 0 & 0 & 0 & 0 & 0 & 0 & 0 & 0 & 0 & 0 & 0 \\
239& 0 & 1 & 0 & 0 & 0 & 0 & 0 & 0 & 0 & 0 & 0 & 0 & 0 & 0 & 0 & 0 & 0 & 0 & 0 & 0 & 0 \\
359& 0 & 1 & 0 & 0 & 0 & 0 & 0 & 0 & 0 & 0 & 0 & 1 & 0 & 1 & 0 & 0 & 0 & 0 & 0 & 0 & 0 \\
479& 0 & 3 & 0 & 1 & 0 & 1 & 2 & 0 & 0 & 1 & 1 & 2 & 0 & 2 & 1 & 1 & 1 & 1 & 1 & 1 & 1 \\
599& 1 & 6 & 1 & 1 & 1 & 4 & 3 & 3 & 2 & 3 & 2 & 8 & 3 & 8 & 6 & 7 & 6 & 8 & 7 & 8 & 9 \\
719& 6 & 17 & 6 & 7 & 7 & 12 & 15 & 12 & 11 & 15 & 17 & 27 & 20 & 34 & 33 & 37 & 39 & 38 & 40 & 42 & 43 \\
839& 23 & 41 & 22 & 24 & 33 & 52 & 52 & 61 & 59 & 71 & 69 & 110 & 94 & 143 & 137 & 167 & 167 & 188 & 187 & 192 & 195
   \\
959& 101 & 134 & 102 & 106 & 124 & 186 & 210 & 224 & 243 & 281 & 294 & 419 & 396 & 561 & 578 & 683 & 705 & 749 & 765
   & 781 & 788 \\
1079& 361 & 418 & 360 & 368 & 469 & 698 & 740 & 866 & 929 & 1058 & 1070 & 1536 & 1485 & 2079 & 2130 & 2564 & 2608 &
   2860 & 2883 & 2931 & 2960 \\
1199& 1291 & 1384 & 1294 & 1312 & 1633 & 2362 & 2612 & 2968 & 3257 & 3694 & 3805 & 5297 & 5233 & 7206 & 7487 & 8960 &
   9195 & 9942 & 10084 & 10238 & 10328 \\
\bottomrule
\end{tabular}
\end{small}
\end{center}

\end{sidewaystable}

\begin{sidewaystable}
\begin{center}
\begin{small}
\caption{$Co_3$ decompositions for $f_7(\tau)$.}
\begin{tabular}{c@{ }|@{\;}r@{ }r@{ }r@{ }r@{ }r@{ }r@{ }r@{ }r@{ }r@{ }r@{ }r@{ }r@{ }r@{ }r@{ }r@{ }r@{ }r@{ }r@{ }r@{ }r@{ }r@{ }r@{ }r@{ }r@{ }r@{ }r@{ }r@{ }r@{ }r@{ }r@{ }r}\toprule
$q^{n/120}$   &$\chi_1$   &$\chi_2$   &$\chi_3$   &$\chi_4$   &$\chi_5$   &$\chi_6$   &$\chi_7$   &$\chi_8$   &$\chi_9$   &$\chi_{10}$   &$\chi_{11}$   &$\chi_{12}$   &$\chi_{13}$  &$\chi_{14}$   &$\chi_{15}$   &$\chi_{16}$   &$\chi_{17}$   &$\chi_{18}$   &$\chi_{19}$   &$\chi_{20}$   &$\chi_{21}$    \\
\midrule
-49& 1 & 0 & 0 & 0 & 0 & 0 & 0 & 0 & 0 & 0 & 0 & 0 & 0 & 0 & 0 & 0 & 0 & 0 & 0 & 0 & 0 \\
71& 0 & 0 & 1 & 0 & 0 & 0 & 0 & 0 & 0 & 0 & 0 & 0 & 0 & 0 & 0 & 0 & 0 & 0 & 0 & 0 & 0 \\
191&  0 & 0 & 0 & 0 & 1 & 0 & 0 & 0 & 0 & 0 & 0 & 0 & 0 & 1 & 0 & 0 & 0 & 0 & 0 & 0 & 0 \\
311& 0 & 0 & 1 & 0 & 0 & 0 & 0 & 0 & 0 & 0 & 0 & 0 & 0 & 0 & 0 & 0 & 0 & 0 & 0 & 0 & 0 \\
431& 1 & 0 & 0 & 0 & 1 & 0 & 0 & 1 & 0 & 0 & 0 & 0 & 0 & 1 & 0 & 0 & 0 & 0 & 0 & 1 & 0 \\
551& 0 & 0 & 2 & 0 & 0 & 0 & 0 & 1 & 0 & 0 & 0 & 1 & 1 & 1 & 0 & 0 & 0 & 0 & 0 & 0 & 1 \\
671& 1 & 0 & 1 & 0 & 2 & 0 & 0 & 4 & 0 & 0 & 0 & 2 & 1 & 4 & 0 & 0 & 0 & 1 & 1 & 5 & 2 \\
791& 0 & 1 & 5 & 1 & 1 & 0 & 0 & 5 & 2 & 0 & 0 & 6 & 4 & 5 & 0 & 4 & 4 & 5 & 5 & 5 & 10 \\
911& 2 & 1 & 4 & 1 & 6 & 2 & 2 & 12 & 2 & 4 & 4 & 12 & 9 & 22 & 11 & 7 & 7 & 20 & 20 & 35 & 30 \\
1031& 0 & 3 & 13 & 4 & 4 & 2 & 2 & 22 & 12 & 12 & 12 & 30 & 30 & 37 & 30 & 47 & 47 & 88 & 88 & 95 & 122 \\
1151& 3 & 4 & 14 & 6 & 17 & 15 & 15 & 59 & 33 & 52 & 52 & 82 & 93 & 143 & 139 & 136 & 136 & 309 & 309 & 386 & 401 \\
\bottomrule
\end{tabular}

\vspace{25pt}
\begin{tabular}{c@{ }|@{\;}r@{ }r@{ }r@{ }r@{ }r@{ }r@{ }r@{ }r@{ }r@{ }r@{ }r@{ }r@{ }r@{ }r@{ }r@{ }r@{ }r@{ }r@{ }r@{ }r@{ }r@{ }r@{ }r@{ }r@{ }r@{ }r@{ }r@{ }r@{ }r@{ }r@{ }r}\toprule
$q^{n/120}$     &$\chi_{22}$   &$\chi_{23}$   &$\chi_{24}$   &$\chi_{25}$   &$\chi_{26}$ &$\chi_{27}$   &$\chi_{28}$   &$\chi_{29}$   &$\chi_{30}$   &$\chi_{31}$  &$\chi_{32}$   &$\chi_{33}$  &$\chi_{34}$   &$\chi_{35}$   &$\chi_{36}$   &$\chi_{37}$   &$\chi_{38}$   &$\chi_{39}$   &$\chi_{40}$   &$\chi_{41}$ &$\chi_{42}$  \\
\midrule
-49& 0 & 0 & 0 & 0 & 0 & 0 & 0 & 0 & 0 & 0 & 0 & 0 & 0 & 0 & 0 & 0 & 0 & 0 & 0 & 0 & 0 \\
71& 0 & 0 & 0 & 0 & 0 & 0 & 0 & 0 & 0 & 0 & 0 & 0 & 0 & 0 & 0 & 0 & 0 & 0 & 0 & 0 & 0 \\
191& 0 & 0 & 0 & 0 & 0 & 0 & 0 & 0 & 0 & 0 & 0 & 0 & 0 & 0 & 0 & 0 & 0 & 0 & 0 & 0 & 0 \\
311& 0 & 1 & 0 & 0 & 0 & 0 & 1 & 0 & 0 & 0 & 0 & 0 & 0 & 0 & 0 & 0 & 0 & 0 & 0 & 0 & 0 \\
431& 0 & 1 & 0 & 0 & 1 & 1 & 0 & 1 & 0 & 0 & 0 & 1 & 0 & 1 & 0 & 0 & 0 & 0 & 0 & 0 & 1 \\
551& 1 & 4 & 1 & 1 & 0 & 1 & 2 & 0 & 0 & 0 & 1 & 2 & 1 & 3 & 3 & 2 & 3 & 1 & 2 & 3 & 2 \\
671& 1 & 5 & 1 & 1 & 4 & 6 & 4 & 6 & 4 & 6 & 4 & 10 & 5 & 12 & 8 & 12 & 10 & 15 & 13 & 14 & 15 \\
791& 10 & 20 & 11 & 12 & 10 & 16 & 23 & 16 & 19 & 24 & 29 & 38 & 33 & 48 & 51 & 57 & 62 & 59 & 63 & 65 & 66 \\
911& 32 & 48 & 31 & 32 & 45 & 75 & 66 & 89 & 87 & 99 & 94 & 154 & 137 & 205 & 200 & 244 & 240 & 277 & 273 & 280 &
   284 \\
1031& 141 & 169 & 143 & 146 & 173 & 244 & 293 & 303 & 335 & 389 & 415 & 565 & 550 & 763 & 797 & 941 & 979 & 1030 &
   1056 & 1074 & 1083 \\
1151& 473 & 521 & 469 & 477 & 621 & 922 & 961 & 1157 & 1236 & 1399 & 1403 & 2016 & 1965 & 2733 & 2797 & 3382 & 3429 &
   3790 & 3811 & 3871 & 3906 \\
\bottomrule
\end{tabular}
\end{small}
\end{center}

\end{sidewaystable}

\begin{sidewaystable}
\begin{center}
\begin{small}
\caption{$Co_2$ decompositions for $f_1(\tau)$.}
\begin{tabular}{c@{ }|@{\;}r@{ }r@{ }r@{ }r@{ }r@{ }r@{ }r@{ }r@{ }r@{ }r@{ }r@{ }r@{ }r@{ }r@{ }r@{ }r@{ }r@{ }r@{ }r@{ }r@{ }r@{ }r@{ }r@{ }r@{ }r@{ }r@{ }r@{ }r@{ }r@{ }r@{ }r}\toprule
$q^{n/120}$   &$\chi_1$   &$\chi_2$   &$\chi_3$   &$\chi_4$   &$\chi_5$   &$\chi_6$   &$\chi_7$   &$\chi_8$   &$\chi_9$   &$\chi_{10}$   &$\chi_{11}$   &$\chi_{12}$   &$\chi_{13}$  &$\chi_{14}$   &$\chi_{15}$   &$\chi_{16}$   &$\chi_{17}$   &$\chi_{18}$   &$\chi_{19}$   &$\chi_{20}$       \\
\midrule
-1&-1 & 0 & 0 & 0 & 0 & 0 & 0 & 0 & 0 & 0 & 0 & 0 & 0 & 0 & 0 & 0 & 0 & 0 & 0 & 0 \\
119& 0 & 0 & 0 & 0 & 1 & 0 & 0 & 0 & 0 & 0 & 0 & 0 & 0 & 0 & 0 & 0 & 0 & 0 & 0 & 0 \\
239& 0 & 0 & 0 & 0 & 0 & 0 & 0 & 1 & 0 & 0 & 0 & 0 & 0 & 0 & 0 & 1 & 0 & 0 & 0 & 0 \\
359& 0 & 0 & 0 & 0 & 1 & 0 & 0 & 1 & 1 & 0 & 0 & 0 & 0 & 0 & 1 & 1 & 0 & 0 & 0 & 0 \\
479& 0 & 1 & 1 & 0 & 2 & 0 & 1 & 1 & 1 & 0 & 0 & 0 & 0 & 0 & 0 & 3 & 0 & 0 & 1 & 0 \\
599& 0 & 1 & 1 & 1 & 4 & 0 & 0 & 3 & 4 & 0 & 0 & 0 & 0 & 0 & 3 & 6 & 0 & 0 & 1 & 0 \\
719& 0 & 2 & 4 & 1 & 6 & 0 & 1 & 6 & 5 & 0 & 0 & 0 & 0 & 0 & 4 & 15 & 0 & 0 & 3 & 0 \\
839& 1 & 3 & 5 & 4 & 14 & 1 & 2 & 11 & 14 & 0 & 0 & 0 & 0 & 1 & 13 & 28 & 1 & 0 & 6 & 1 \\
959& 0 & 5 & 11 & 5 & 23 & 0 & 4 & 22 & 25 & 0 & 0 & 0 & 0 & 0 & 22 & 62 & 2 & 0 & 17 & 4 \\
1079& 2 & 7 & 17 & 12 & 47 & 2 & 7 & 42 & 54 & 0 & 0 & 0 & 0 & 2 & 56 & 121 & 11 & 0 & 34 & 11 \\
1199& 2 & 13 & 36 & 17 & 82 & 4 & 16 & 80 & 98 & 2 & 2 & 0 & 0 & 6 & 106 & 255 & 26 & 5 & 82 & 34 \\
 \bottomrule
\end{tabular}

\vspace{25pt}
\begin{tabular}{c@{ }|@{\;}r@{ }r@{ }r@{ }r@{ }r@{ }r@{ }r@{ }r@{ }r@{ }r@{ }r@{ }r@{ }r@{ }r@{ }r@{ }r@{ }r@{ }r@{ }r@{ }r@{ }r@{ }r@{ }r@{ }r@{ }r@{ }r@{ }r@{ }r@{ }r@{ }r@{ }r}\toprule
$q^{n/120}$  &$\chi_{21}$   &$\chi_{22}$   &$\chi_{23}$   &$\chi_{24}$   &$\chi_{25}$   &$\chi_{26}$ &$\chi_{27}$   &$\chi_{28}$   &$\chi_{29}$   &$\chi_{30}$   &$\chi_{31}$  &$\chi_{32}$   &$\chi_{33}$  &$\chi_{34}$   &$\chi_{35}$   &$\chi_{36}$   &$\chi_{37}$   &$\chi_{38}$   &$\chi_{39}$   &$\chi_{40}$    \\
\midrule
-1&0 & 0 & 0 & 0 & 0 & 0 & 0 & 0 & 0 & 0 & 0 & 0 & 0 & 0 & 0 & 0 & 0 & 0 & 0 & 0 \\
119& 0 & 0 & 0 & 0 & 0 & 0 & 0 & 0 & 0 & 0 & 0 & 0 & 0 & 0 & 0 & 0 & 0 & 0 & 0 & 0 \\
239& 0 & 0 & 0 & 0 & 0 & 0 & 0 & 0 & 0 & 0 & 0 & 0 & 0 & 0 & 0 & 0 & 0 & 0 & 0 & 0 \\
359& 0 & 0 & 0 & 0 & 1 & 1 & 0 & 0 & 0 & 0 & 0 & 0 & 0 & 0 & 0 & 0 & 0 & 0 & 0 & 0 \\
479& 2 & 0 & 0 & 0 & 2 & 2 & 0 & 0 & 0 & 0 & 0 & 0 & 0 & 0 & 0 & 1 & 1 & 0 & 0 & 0 \\
599& 3 & 0 & 0 & 1 & 5 & 5 & 0 & 0 & 0 & 0 & 0 & 0 & 0 & 0 & 1 & 4 & 2 & 0 & 0 & 0 \\
719& 8 & 0 & 0 & 3 & 12 & 13 & 0 & 0 & 0 & 0 & 0 & 0 & 0 & 1 & 4 & 9 & 7 & 0 & 0 & 0 \\
839& 17 & 0 & 0 & 9 & 28 & 32 & 1 & 1 & 0 & 2 & 0 & 0 & 1 & 6 & 10 & 25 & 17 & 2 & 2 & 2 \\
959& 43 & 1 & 1 & 18 & 64 & 72 & 3 & 5 & 4 & 7 & 1 & 1 & 7 & 15 & 32 & 59 & 51 & 12 & 10 & 10 \\
1079& 88 & 7 & 7 & 50 & 144 & 167 & 21 & 20 & 20 & 30 & 13 & 13 & 28 & 53 & 86 & 155 & 126 & 46 & 49 & 49 \\
1199& 203 & 32 & 32 & 120 & 322 & 378 & 73 & 82 & 83 & 103 & 67 & 67 & 102 & 155 & 247 & 375 & 348 & 175 & 171 & 177
   \\
\bottomrule
\end{tabular}
\vspace{25pt}

\begin{tabular}{c@{ }|@{\;}r@{ }r@{ }r@{ }r@{ }r@{ }r@{ }r@{ }r@{ }r@{ }r@{ }r@{ }r@{ }r@{ }r@{ }r@{ }r@{ }r@{ }r@{ }r@{ }r@{ }r@{ }r@{ }r@{ }r@{ }r@{ }r@{ }r@{ }r@{ }r@{ }r@{ }r}\toprule
$q^{n/120}$  &$\chi_{41}$   &$\chi_{42}$   &$\chi_{43}$   &$\chi_{44}$   &$\chi_{45}$   &$\chi_{46}$ &$\chi_{47}$   &$\chi_{48}$   &$\chi_{49}$   &$\chi_{50}$   &$\chi_{51}$  &$\chi_{52}$   &$\chi_{53}$  &$\chi_{54}$   &$\chi_{55}$   &$\chi_{56}$   &$\chi_{57}$   &$\chi_{58}$   &$\chi_{59}$   &$\chi_{60}$    \\
\midrule
-1& 0 & 0 & 0 & 0 & 0 & 0 & 0 & 0 & 0 & 0 & 0 & 0 & 0 & 0 & 0 & 0 & 0 & 0 & 0 & 0 \\
119& 0 & 0 & 0 & 0 & 0 & 0 & 0 & 0 & 0 & 0 & 0 & 0 & 0 & 0 & 0 & 0 & 0 & 0 & 0 & 0 \\
239& 0 & 0 & 0 & 0 & 0 & 0 & 0 & 0 & 0 & 0 & 0 & 0 & 0 & 0 & 0 & 0 & 0 & 0 & 0 & 0 \\
359& 0 & 0 & 0 & 0 & 0 & 0 & 0 & 0 & 0 & 0 & 0 & 0 & 0 & 0 & 0 & 0 & 0 & 0 & 0 & 0 \\
479& 0 & 0 & 0 & 1 & 0 & 1 & 0 & 0 & 0 & 0 & 0 & 0 & 0 & 0 & 0 & 0 & 0 & 0 & 0 & 0 \\
599& 0 & 0 & 0 & 3 & 0 & 3 & 0 & 0 & 0 & 2 & 0 & 0 & 0 & 0 & 2 & 0 & 0 & 0 & 1 & 0 \\
719& 0 & 2 & 1 & 7 & 0 & 11 & 0 & 2 & 0 & 6 & 0 & 2 & 2 & 3 & 9 & 0 & 2 & 3 & 4 & 1 \\
839& 0 & 6 & 6 & 23 & 2 & 28 & 4 & 6 & 2 & 25 & 5 & 10 & 10 & 15 & 32 & 7 & 11 & 17 & 18 & 12 \\
959& 5 & 26 & 20 & 60 & 12 & 82 & 17 & 30 & 17 & 77 & 31 & 52 & 52 & 63 & 108 & 42 & 59 & 70 & 75 & 59 \\
1079& 30 & 76 & 74 & 174 & 61 & 213 & 81 & 96 & 80 & 256 & 138 & 208 & 209 & 245 & 352 & 201 & 236 & 273 & 281 & 247
   \\
1199& 144 & 268 & 243 & 464 & 239 & 595 & 286 & 348 & 323 & 777 & 531 & 777 & 775 & 863 & 1122 & 778 & 887 & 958 &
   983 & 910 \\
\bottomrule
\end{tabular}

\end{small}
\end{center}

\end{sidewaystable}

\begin{sidewaystable}
\begin{center}
\begin{small}
\caption{$Co_2$ decompositions for $f_7(\tau)$.}
\begin{tabular}{c@{ }|@{\;}r@{ }r@{ }r@{ }r@{ }r@{ }r@{ }r@{ }r@{ }r@{ }r@{ }r@{ }r@{ }r@{ }r@{ }r@{ }r@{ }r@{ }r@{ }r@{ }r@{ }r@{ }r@{ }r@{ }r@{ }r@{ }r@{ }r@{ }r@{ }r@{ }r@{ }r}\toprule
$q^{n/120}$   &$\chi_1$   &$\chi_2$   &$\chi_3$   &$\chi_4$   &$\chi_5$   &$\chi_6$   &$\chi_7$   &$\chi_8$   &$\chi_9$   &$\chi_{10}$   &$\chi_{11}$   &$\chi_{12}$   &$\chi_{13}$  &$\chi_{14}$   &$\chi_{15}$   &$\chi_{16}$   &$\chi_{17}$   &$\chi_{18}$   &$\chi_{19}$   &$\chi_{20}$       \\
\midrule
-49&1 & 0 & 0 & 0 & 0 & 0 & 0 & 0 & 0 & 0 & 0 & 0 & 0 & 0 & 0 & 0 & 0 & 0 & 0 & 0 \\
71& 0 & 0 & 1 & 0 & 0 & 0 & 0 & 0 & 0 & 0 & 0 & 0 & 0 & 0 & 0 & 0 & 0 & 0 & 0 & 0 \\
191& 0 & 0 & 0 & 1 & 0 & 0 & 0 & 0 & 1 & 0 & 0 & 0 & 0 & 0 & 0 & 0 & 0 & 0 & 0 & 0 \\
311& 0 & 0 & 1 & 0 & 0 & 0 & 0 & 0 & 0 & 0 & 0 & 0 & 0 & 0 & 0 & 1 & 0 & 0 & 0 & 0 \\
431& 1 & 0 & 0 & 1 & 1 & 0 & 0 & 0 & 1 & 0 & 0 & 0 & 0 & 0 & 1 & 1 & 0 & 0 & 0 & 0 \\
551& 0 & 0 & 2 & 0 & 1 & 0 & 0 & 1 & 1 & 0 & 0 & 0 & 0 & 0 & 0 & 4 & 0 & 0 & 1 & 0 \\
671& 1 & 0 & 1 & 2 & 4 & 0 & 0 & 2 & 4 & 0 & 0 & 0 & 0 & 0 & 4 & 5 & 0 & 0 & 1 & 0 \\
791& 0 & 1 & 5 & 1 & 5 & 0 & 1 & 5 & 5 & 0 & 0 & 0 & 0 & 0 & 3 & 15 & 0 & 0 & 4 & 1 \\
911& 2 & 1 & 4 & 5 & 12 & 1 & 1 & 10 & 15 & 0 & 0 & 0 & 0 & 1 & 14 & 26 & 2 & 0 & 6 & 1 \\
1031& 0 & 3 & 13 & 4 & 20 & 0 & 4 & 19 & 21 & 0 & 0 & 0 & 0 & 0 & 19 & 61 & 2 & 0 & 17 & 5 \\
1151& 3 & 4 & 14 & 13 & 43 & 3 & 6 & 37 & 52 & 0 & 0 & 0 & 0 & 3 & 59 & 112 & 14 & 1 & 33 & 13 \\\bottomrule
\end{tabular}
\vspace{25pt}

\begin{tabular}{c@{ }|@{\;}r@{ }r@{ }r@{ }r@{ }r@{ }r@{ }r@{ }r@{ }r@{ }r@{ }r@{ }r@{ }r@{ }r@{ }r@{ }r@{ }r@{ }r@{ }r@{ }r@{ }r@{ }r@{ }r@{ }r@{ }r@{ }r@{ }r@{ }r@{ }r@{ }r@{ }r}\toprule
$q^{n/120}$  &$\chi_{21}$   &$\chi_{22}$   &$\chi_{23}$   &$\chi_{24}$   &$\chi_{25}$   &$\chi_{26}$ &$\chi_{27}$   &$\chi_{28}$   &$\chi_{29}$   &$\chi_{30}$   &$\chi_{31}$  &$\chi_{32}$   &$\chi_{33}$  &$\chi_{34}$   &$\chi_{35}$   &$\chi_{36}$   &$\chi_{37}$   &$\chi_{38}$   &$\chi_{39}$   &$\chi_{40}$    \\
\midrule
-49&0 & 0 & 0 & 0 & 0 & 0 & 0 & 0 & 0 & 0 & 0 & 0 & 0 & 0 & 0 & 0 & 0 & 0 & 0 & 0 \\
 71&0 & 0 & 0 & 0 & 0 & 0 & 0 & 0 & 0 & 0 & 0 & 0 & 0 & 0 & 0 & 0 & 0 & 0 & 0 & 0 \\
191& 0 & 0 & 0 & 0 & 0 & 0 & 0 & 0 & 0 & 0 & 0 & 0 & 0 & 0 & 0 & 0 & 0 & 0 & 0 & 0 \\
311& 1 & 0 & 0 & 0 & 0 & 0 & 0 & 0 & 0 & 0 & 0 & 0 & 0 & 0 & 0 & 0 & 0 & 0 & 0 & 0 \\
431& 0 & 0 & 0 & 1 & 1 & 1 & 0 & 0 & 0 & 0 & 0 & 0 & 0 & 0 & 0 & 1 & 0 & 0 & 0 & 0 \\
551& 2 & 0 & 0 & 0 & 2 & 2 & 0 & 0 & 0 & 0 & 0 & 0 & 0 & 0 & 1 & 1 & 1 & 0 & 0 & 0 \\
671& 3 & 0 & 0 & 2 & 5 & 6 & 0 & 0 & 0 & 0 & 0 & 0 & 0 & 1 & 1 & 5 & 2 & 0 & 0 & 0 \\
791& 10 & 0 & 0 & 2 & 12 & 13 & 0 & 1 & 0 & 0 & 0 & 0 & 0 & 1 & 5 & 9 & 9 & 0 & 0 & 0 \\
911& 15 & 0 & 0 & 10 & 29 & 33 & 3 & 1 & 1 & 3 & 0 & 0 & 2 & 7 & 12 & 29 & 18 & 3 & 5 & 5 \\
1031& 45 & 2 & 2 & 19 & 64 & 73 & 5 & 7 & 7 & 9 & 4 & 4 & 9 & 18 & 38 & 60 & 58 & 17 & 14 & 14 \\
1151& 85 & 11 & 11 & 57 & 145 & 171 & 32 & 30 & 29 & 42 & 22 & 22 & 40 & 68 & 96 & 169 & 137 & 64 & 70 & 70 \\\bottomrule
\end{tabular}
\vspace{25pt}

\begin{tabular}{c@{ }|@{\;}r@{ }r@{ }r@{ }r@{ }r@{ }r@{ }r@{ }r@{ }r@{ }r@{ }r@{ }r@{ }r@{ }r@{ }r@{ }r@{ }r@{ }r@{ }r@{ }r@{ }r@{ }r@{ }r@{ }r@{ }r@{ }r@{ }r@{ }r@{ }r@{ }r@{ }r}\toprule
$q^{n/120}$  &$\chi_{41}$   &$\chi_{42}$   &$\chi_{43}$   &$\chi_{44}$   &$\chi_{45}$   &$\chi_{46}$ &$\chi_{47}$   &$\chi_{48}$   &$\chi_{49}$   &$\chi_{50}$   &$\chi_{51}$  &$\chi_{52}$   &$\chi_{53}$  &$\chi_{54}$   &$\chi_{55}$   &$\chi_{56}$   &$\chi_{57}$   &$\chi_{58}$   &$\chi_{59}$   &$\chi_{60}$    \\
\midrule
-49&0 & 0 & 0 & 0 & 0 & 0 & 0 & 0 & 0 & 0 & 0 & 0 & 0 & 0 & 0 & 0 & 0 & 0 & 0 & 0 \\
71& 0 & 0 & 0 & 0 & 0 & 0 & 0 & 0 & 0 & 0 & 0 & 0 & 0 & 0 & 0 & 0 & 0 & 0 & 0 & 0 \\
191& 0 & 0 & 0 & 0 & 0 & 0 & 0 & 0 & 0 & 0 & 0 & 0 & 0 & 0 & 0 & 0 & 0 & 0 & 0 & 0 \\
311& 0 & 0 & 0 & 0 & 0 & 0 & 0 & 0 & 0 & 0 & 0 & 0 & 0 & 0 & 0 & 0 & 0 & 0 & 0 & 0 \\
431& 0 & 0 & 0 & 0 & 0 & 0 & 0 & 0 & 0 & 0 & 0 & 0 & 0 & 0 & 0 & 0 & 0 & 0 & 0 & 0 \\
551& 0 & 0 & 0 & 0 & 0 & 2 & 0 & 0 & 0 & 0 & 0 & 0 & 0 & 0 & 1 & 0 & 0 & 0 & 0 & 0 \\
671& 0 & 0 & 1 & 4 & 0 & 3 & 0 & 0 & 0 & 3 & 0 & 0 & 0 & 1 & 3 & 0 & 0 & 1 & 1 & 1 \\
791& 0 & 4 & 2 & 8 & 0 & 13 & 0 & 4 & 1 & 8 & 1 & 4 & 4 & 5 & 12 & 1 & 4 & 5 & 6 & 3 \\
911& 1 & 6 & 8 & 27 & 4 & 31 & 8 & 7 & 5 & 33 & 10 & 16 & 16 & 22 & 41 & 15 & 18 & 25 & 27 & 20 \\
1031& 11 & 36 & 27 & 66 & 19 & 94 & 24 & 43 & 27 & 93 & 48 & 77 & 76 & 90 & 137 & 65 & 89 & 98 & 104 & 86 \\
1151& 46 & 92 & 96 & 199 & 90 & 236 & 115 & 120 & 115 & 319 & 195 & 282 & 284 & 324 & 436 & 290 & 319 & 363 & 371 &
   340 \\
\bottomrule
\end{tabular}

\end{small}
\end{center}

\end{sidewaystable}

\clearpage


\end{document}